\documentclass[prd,amsmath,amssymb,aps,twocolumn,nofootinbib]{revtex4-1}

\usepackage{subcaption}
\usepackage[dvipsnames]{xcolor}
\usepackage{graphicx}
\usepackage{todonotes}
\usepackage{comment}
\usepackage[colorlinks=true,citecolor=blue]{hyperref}
\usepackage{ulem}
\usepackage{soul}
\setstcolor{Green}
\usepackage{cancel}
\usepackage{mathrsfs}
\DeclareMathAlphabet{\mathcalligra}{T1}{calligra}{m}{n}
\usepackage[utf8]{inputenc}

\newcommand{\subE}{\textrm{\tiny{E}}}

\newcommand{\be}{\begin{equation} }
	\newcommand{\ee}{\end{equation}}
\newcommand{\bes}{\begin{equation*} }
	\newcommand{\ees}{\end{equation*}}
\newcommand{\bea}{\begin{eqnarray} }
	\newcommand{\eea}{\end{eqnarray}}
\newcommand{\beas}{\begin{eqnarray*} }
	\newcommand{\eeas}{\end{eqnarray*}}
\newcommand{\ba}{\begin{align} }
	\newcommand{\ea}{\end{align} }
\newcommand{\bas}{\begin{align*} }
	\newcommand{\eas}{\end{align*} }

\usepackage{caption}
\newcommand{\jar}[1]{\slshape{\color{violet}}}

\begin{document}
	
	\title{The renormalized stress-energy tensor for scalar fields in the Boulware state with applications to extremal black holes}
	\author{Julio Arrechea}
	\email{julio.arrechea@sissa.it}
	\affiliation{IFPU, Institute for Fundamental Physics of the Universe, via Beirut 2, 34014 Trieste, Italy}
    \affiliation{SISSA, International School for Advanced Studies,
     via Bonomea 265, 34136 Trieste, Italy}
    \affiliation{INFN Sezione di Trieste,
     via Valerio 2, 34127 Trieste, Italy} 
	
	\author{Cormac Breen}
	\email{cormac.breen@tudublin.ie}
	\affiliation{School of Mathematics and Statistics, Technological University Dublin, Grangegorman, Dublin 7, Ireland}

\author{Adrian Ottewill}
	\email{adrian.ottewill@ucd.ie}
	\affiliation{School of Mathematics and Statistics, University College Dublin, Belfield, Dublin 4, Ireland}

 \author{Lorenzo Pisani}
	\email{lorenzo.pisani2@mail.dcu.ie}
	\affiliation{
		Center for Astrophysics and Relativity, School of Mathematical Sciences, Dublin City University, Glasnevin, Dublin 9, Ireland}
	
\author{Peter Taylor}
	\email{peter.taylor@dcu.ie}
	\affiliation{
		Center for Astrophysics and Relativity, School of Mathematical Sciences, Dublin City University, Glasnevin, Dublin 9, Ireland}
		
	\date{\today}
	\begin{abstract}
	We provide a mode-sum prescription to directly compute the renormalized stress-energy tensor (RSET) for scalar fields in the Boulware vacuum. The method generalizes the recently developed extended coordinate method which was previously only applicable to Hartle-Hawking states. We exhibit the accuracy and efficiency of the method by calculating the RSET in sub-extremal and extremal Reissner-Nordström spacetimes. We find numerical evidence for the regularity of the RSET at the extremal horizon regardless of the field mass and its coupling. We employ our numerical results of the RSET to source the semi-classical Einstein equations, demonstrating that if the RSET is considered as a static perturbation, it will either de-extremalize the black hole, or convert it into a horizonless object.
	\end{abstract}
	\maketitle
	
 \section{Introduction}
 In the semiclassical approximation to quantum gravity, one replaces the stress-energy tensor in the gravitational field equations with the corresponding expectation value for the quantized matter fields, including ostensibly quantized graviton fields considered as linear perturbations about a background geometry. The gravitational field equations in this framework are
\begin{align}\label{eq:semieqs}
		 G_{\mu\nu}+\Lambda\,g_{\mu\nu}+\alpha_{1} H^{(1)}_{\mu\nu}+\alpha_{2}\,H^{(2)}_{\mu\nu}=8\pi\,\left(T^{\textrm{(cl)}}_{\mu\nu}+\langle\hat{T}_{\mu\nu}\rangle_{\textrm{ren}}\right),
	\end{align}
where $g_{\mu\nu}$ is the metric of spacetime, $\Lambda$ is the cosmological constant, $G_{\mu\nu}$ is the Einstein tensor and $\langle\hat{T}_{\mu\nu}\rangle_{\textrm{ren}}$ is the renormalized expectation value of the stress-energy tensor (RSET) of the quantum fields in some quantum state. The tensors $H^{(1)}_{\mu\nu}$, $H^{(2)}_{\mu\nu}$ are geometrical and include terms that are quadratic in the curvature; their inclusion arises through the point-splitting regularization process that yields $\langle\hat{T}_{\mu\nu}\rangle_{\textrm{ren}}$. This regularization process allows for the calculation of the finite physical RSET with an infinite renormalization of the constants $\Lambda$, $\alpha_{1}$ and $\alpha_{2}$. It is difficult to conceive how to practically solve Eq.~(\ref{eq:semieqs}) in a general self-consistent way since the right-hand side depends on the unknown solution $g_{\mu\nu}$, and yet to compute the stress-energy tensor explicitly requires that we know the geometry of spacetime \textit{a priori}. It is only by considering highly symmetric spacetimes~\cite{Anderson1983} or by subjecting the RSET to additional approximations~\cite{Parker:1993dk,ParentaniPiran1994,Gao:2023obj} that the backreaction problem becomes tractable.
In black hole spacetimes, one usually proceeds by a perturbative expansion of the metric about a background classical solution. This leads to the reduced order semiclassical equations which describe, to first order in $\hbar$, the quantum backreaction induced by a quantum field on the classical geometry~\cite{York1984,FlanaganWald1996,taylorbreen:2020}. In these equations, the source term is the renormalized expectation value for a quantum field in a given quantum state on a fixed classical background spacetime.

It turns out that in most situations of interest, even the calculation of the renormalized expectation value of the stress-energy tensor (RSET) for a particular quantum state on a fixed background is a technically challenging endeavour. For black hole spacetimes, there are three main approaches to this calculation, the Candelas-Howard approach \cite{CandelasHoward:1984} and its extensions (see for example \cite{AHSPRL1993,AHS1995, BreenOttewill2012}), the so-called ``pragmatic mode-sum prescription" \cite{LeviOri:2015, LeviOri:2016,LeviOri:2016b} and the method developed by authors of this paper called the ``extended coordinate method" \cite{taylorbreen:2016,taylorbreen:2017, taylorbreenottewill:2021,Arrechea:2023}. While all these methods have different approaches and advantages, one drawback of both the original Candelas-Howard approach and the extended coordinate approach is that they were developed specifically for the field in the Hartle-Hawking state \cite{HartleHawking:1976}. Working with this state offers a computational advantage since one can employ Euclidean techniques that enforce a discretization of the frequency spectrum. Anderson, Hiscock and Samuel \cite{AHS1995} generalized the Candelas-Howard method to include a direct method for computing the RSET for a scalar field in the Boulware state \cite{Boulware:1975}. Although the RSET in the Unruh, Boulware and Hartle-Hawking state were computed in Ref.~\cite{Arrechea:2023}, only the Hartle-Hawking state was directly renormalized, the RSET for the other two states were computed by leveraging the fact that the difference between two quantum states is regular. In this paper, we generalize the extended coordinate method to directly compute the RSET for a scalar field in the Boulware state.

A natural question then is: if the RSET for a field in the  Boulware state can be computed by first renormalizing in the Hartle-Hawking state and employing a state subtraction scheme, then why do we need a direct method for renormalizing in the Boulware state? There are several justifications for a direct renormalization prescription for Boulware. First, when employing the indirect method one needs to compute both the Euclidean modes (to do the renormalization in the Hartle-Hawking state) and the Lorentzian modes (to do the state subtraction scheme). Employing a direct method for Boulware requires only one set of modes and hence is computationally less expensive. Second, there are black hole spacetimes that do not admit a Hartle-Hawking state, for example, the extreme Reissner-Nordstrom spacetime, and hence one requires a direct method for computing the RSET in such situations. Finally, moving away from black hole spacetimes, if we consider semiclassical effects in the spacetime of a star, then the natural vacuum state is the Boulware state (there is, in fact, no other reference state). Hence this paper sets the foundation for renormalization in those contexts.

The approach we take is similar to that taken in the extended coordinate method for the Hartle-Hawking state \cite{taylorbreen:2016,taylorbreen:2017, taylorbreenottewill:2021}. We take an arbitrarily high-order expansion of the Hadamard parametrix which encodes the short-distance behaviour of the two-point function, and we rewrite the terms in this expansion as mode-sums of the same form as the Boulware propagator. While the spirit is the same, the mathematical development is significantly trickier because the frequency spectrum is continuous, rather than the discrete frequency spectrum that one obtains in the Euclidean approach to defining the Hartle-Hawking state. As well as presenting a computational challenge, the continuous frequency spectrum also means that the Fourier transform of the terms in the Hadamard parametrix, which is needed to express these as mode-sums, yields functions of the frequency that are not ordinary functions but rather generalized functions. When we treat these mode-sums rigorously as generalized functions, it becomes completely transparent how an infra-red cutoff appears in these $\omega$ integrals and indeed it is completely explicit that the RSET is independent of the choice of this cutoff. A similar cutoff was introduced in Ref.~\cite{AHS1995} but it was introduced by necessity to eliminate an infra-red divergence in the WKB approximation. While the approaches are ultimately equivalent, it is clear here that the source of the cutoff is a consequence of treating rigorously the generalized functions that arise in the Fourier transform of the Hadamard parametrix.

As a consistency check, we contrasted our numerical results with previously known results for the sub-extremal and extremal Reissner-Nordström spacetime~\cite{Arrechea:2023,AndersonEBH} and found excellent agreement. We pay particular attention to the regularity of the RSET at the horizon of the extreme Reissner-Nordström black hole. Numerical evidence suggesting its regularity was found in~\cite{AndersonEBH} for massless fields (see~\cite{Anderson2000,Lowe2001,Anderson2001,Matyjasek:2001yf} for discussions about backreaction effects). A clear advantage of the extended coordinate method over some previous approaches is that it does not invoke the Wentzel–Kramers–Brillouin (WKB) approximation in any step of the calculation, allowing to generate very accurate results for the RSET near the horizon. Our results reinforce the existing numerical evidence of the regularity of the RSET in extremal black holes. Additionally, we provide the first extension of those results to massive fields.

This paper is organized as follows. In Sec.~\ref{sec:Hadamard}, we show how we express the expansion of the Hadamard parametrix as a mode-sum of the same type as the Boulware propagator for a scalar field in a static, spherically symmetric spacetime. In Sec.~\ref{sec:Reg}, we give the explicit expressions for the renormalized expectation values for the vacuum polarization and the stress-energy tensor for a scalar field in the Boulware state. In Sec.~\ref{sec:ReissnerNordstrom}, we apply the new method to compute the RSET for a scalar field on the Reissner-Nordstrom spacetime, including the extremal case where a direct calculation is necessary. In Sec.~\ref{sec:backreaction}, we employ our accurate RSET results to solve the reduced-order semiclassical equations to compute the backreaction on the extremal Reissner-Nordstrom black hole spacetime. Finally, in Sec.~\ref{sec:conclusions}, we conclude with some discussion and future prospects.

\section{Renormalization Prescription in the Boulware State}
 \label{sec:Hadamard}
In this section, we will briefly outline the extended coordinate approach to calculating the RSET for a quantum scalar field in the Boulware quantum state \cite{Boulware:1975}, propagating on a static, spherically symmetric black hole spacetime.  
The Boulware state is vacuum according to a stationary observer far from the black hole. It is useful to treat the Boulware state as a thermal state at zero temperature and to employ Euclidean methods. We consider Euclidean line elements of the form:
\begin{align}
	\label{eq:metric}	ds^{2}=f(r)d\tau^{2}+dr^{2}/f(r)+r^{2}( d\theta^2 + \sin^2 \theta d \phi^2).	\end{align}
If this line element describes a black hole spacetime with event horizon $r=r_{+}$ and surface gravity $\kappa=\tfrac{1}{2}f'(r_{+})$, then the geometry possesses a conical singularity at the horizon unless the Euclidean time is periodic with periodicity $\tau=\tau+2\pi/\kappa$. If we wished to consider a quantum field in the Hartle-Hawking state, this identification would be the appropriate one. Moreover, imposing a periodicity discretizes the frequency spectrum of the field modes, which is a major computational advantage. For computing the renormalization counter-terms, all that matters is the geometry, and so there is no obligation to make any particular identification on the Euclidean time; nevertheless, the renormalized mode-sum is most conveniently derived and computed when we impose the same periodicity on the mode-sum representation of the renormalization counter-terms as we do on the quantum field in the Hartle-Hawking state. In this paper, we are interested in the case where the quantum field is in the Boulware state, which is a zero temperature pure state. We make no periodic identification in Euclidean time which implies that the frequency spectrum of the quantum field (and the mode-sum representation of the renormalization counter-terms) will be continuous and that our geometry retains the conical singularity at the event horizon, at least for non-extremal black holes.
Given that one of the main motivations for adopting Euclidean techniques in the Hartle-Hawking case is the computational advantage obtained by the discrete frequency spectrum, one might be tempted to discard the Euclidean approach for computing quantum expectation values in the Boulware state, where the spectrum remains continuous. However,  
the Euclidean approach is still useful since, unlike the Lorentzian framework,  the Euclidean Green function does not require an `$i\,\epsilon$' prescription to define the bi-distribution along the lightcone (since there is no lightcone in a Euclidean space). In particular, on a Euclidean geometry, the Klein-Gordon equation for the scalar field is elliptic
\begin{align}\label{eq:kg}
	(\Box-\mu^2-\xi\,R)\phi=0,
\end{align}
where $\Box$ is the d'Alembertian operator with respect to the Euclidean metric, $\mu$ is the field mass with dimensions of inverse length, $R$ is the Ricci curvature scalar of the background spacetime and $\xi$ is the coupling strength between the field and the background geometry. The corresponding Euclidean Green function has the following mode-sum representation
(with $r=r'$ for simplicity) on this black hole spacetime,
\begin{align}
	\label{eq:Gmodesum}
	&G(x,x')=\frac{1}{8 \pi^2}\sum_{l=0}^{\infty}(2l+1)P_l(\cos\gamma)\int_{-\infty}^{\infty}d\omega\,e^{i\omega\Delta\tau}g_{\omega l}(r),
\end{align}
where $\Delta x\equiv x'-x\sim\mathcal{O}(\epsilon)$ is the coordinate separation, $\gamma$ is the geodesic distance on the 2-sphere and $P_{l}(z)$ is the Legendre polynomial of the first kind. We have denoted by $g_{\omega l}(r)=p_{\omega l}(r)\,q_{\omega l}(r)/N_{\omega l}$ the one-dimensional radial Green function evaluated at the same radius. The radial modes $p_{\omega l}(r)$, $q_{\omega l}(r)$ are solutions of the homogeneous equation:
\begin{align}
\label{eq:Euclideanradialeqn}
	\bigg[\frac{d}{dr}\Big(r^2f(r)\frac{d}{dr}\Big)-r^{2}\left(\frac{\omega^{2}}{f(r)}+(\mu^{2}+\xi\,R)\right)\nonumber\\
	-l(l+1)\bigg]\chi_{\omega l}(r)=0,
\end{align}
with $p_{\omega l}(r)$ the solution regular at the horizon and $q_{\omega l}(r)$ regular at the outer boundary (usually spatial infinity or the cosmological horizon). The normalization constant is given by
\begin{align}
	N_{\omega l}=-r^{2}f(r)\,\mathcal{W}\{p_{\omega l}(r),q_{\omega l}(r)\},
\end{align}
where $\mathcal{W}\{p,q\}$ denotes the Wronskian of the two solutions.

In the coincidence limit $\Delta x\to 0$ (i.e. $\gamma\to 0$ and $\Delta\tau\to 0$), the mode sum (\ref{eq:Gmodesum}) diverges. To renormalize this mode sum, we will adapt the so-called extended coordinate implementation of Hadamard renormalization developed in Refs.~\cite{taylorbreen:2016,taylorbreen:2017, taylorbreenottewill:2021}. For a quantum state satisfying the Hadamard condition, the local singularity structure of the two-point function when $x$ and $x'$ are sufficiently close is encapsulated in the Hadamard parametrix
\begin{align}
\label{eq:HadamardForm}
    K(x,x')=\frac{1}{8\pi^{2}}\left(\frac{\Delta^{1/2}(x,x')}{\sigma(x,x')}+V(x,x')\log\Big(\frac{2\sigma(x,x')}{\ell^{2}}\Big)\right)
\end{align}
where $\sigma(x,x')$ is Synge's world function which measures half the square of the geodesic distance between $x$ and $x'$, $\Delta^{1/2}(x,x')$ and $V(x,x')$ are symmetric biscalars that are regular in the coincidence limit $x'\to x$. Moreover, these biscalars depend only on the local geometry and the field parameters, not on any global considerations such as the quantum state. The lengthscale $\ell$ here is an arbitrary lengthscale associated with the renormalization ambiguity. It arises since $V(x,x')$ is a solution to the homogeneous wave equation and so one is free to add arbitrary factors of $V(x,x')$ to $K(x,x')$ and it will remain a parametrix for the wave operator. Since $\sigma(x,x')\to 0$ as $x'\to x$, it is clear that $K(x,x')$ is singular at coincidence.

The fact that the mode-sum (\ref{eq:Gmodesum}) does not converge at coincidence is because the two-point function possesses singularities precisely of the type in Eq.~(\ref{eq:HadamardForm}) in the short-distance limit. The goal is to subtract $K(x,x')$ from $G(x,x')$ in such a way that the coincidence limit can be meaningfully taken. The only way to proceed in this approach is to express $K(x,x')$ as a mode-sum of the same type as the mode-sum representation of Eq.~(\ref{eq:Gmodesum}) and then to subtract mode-by-mode. This renormalized mode-sum will converge in the coincidence limit.

Expressing the Hadamard parametrix as a mode-sum is non-trivial. The approach of Refs.~\cite{taylorbreen:2016,taylorbreen:2017} is to expand the parametrix in a judiciously chosen set of extended coordinates. In that context, it was assumed the field was in the Hartle-Hawking state and the extended coordinates were adapted to that state by ensuring the time dependence had the appropriate thermal periodicity. This must be amended here for the Boulware state. We define
\begin{align}  
s^{2}=
 f(r)\,\Delta\tau^{2}+2 r^{2}(1-\cos\gamma),
\end{align}
and then, rather than expand the parametrix in coordinate separation $\Delta x$, we expand in  $s$ and $\Delta \tau$ (we take radial points together $\Delta r=0$). The result of expanding the parametrix up to $\mathcal{O}(\epsilon^{2m}\log\epsilon)$ in these new extended coordinates is
\begin{align}
\label{eq:hadamard_mode_sum_1}
    K(x,x')&=\frac{1}{8\pi^{2}}\Bigg(\sum_{a=0}^{m}\sum_{b=0}^{a}\mathcal{D}_{ab}^{(\textrm{r})}(r)\frac{\Delta\tau^{2a+2b}}{s^{2b+2}}\nonumber\\
&+\sum_{a=1}^{m}\sum_{b=1}^{a}\mathcal{D}_{ab}^{(\textrm{p})}(r)\Delta\tau^{2a-2b}s^{2b-2}\nonumber\\
&+\sum_{a=1}^{m-1}\sum_{b=0}^{a-1}\mathcal{T}^{(\textrm{r})}_{ab}(r)\frac{\Delta\tau^{2a+2b+2}}{s^{2b+2}} \nonumber \\
& +\sum_{a=0}^{m-1}\sum_{b=0}^{a}\mathcal{T}_{ab}^{(\textrm{l})}(r)s^{2a-2b}\Delta\tau^{2b}\log\left(\frac{s^{2}}{\ell^{2}}\right)\nonumber\\
&+\sum_{a=1}^{m-1}\sum_{b=0}^{a}\mathcal{T}_{ab}^{(\textrm{p})}(r)s^{2a-2b}\Delta\tau^{2b}\Bigg)+\mathcal{O}(\epsilon^{2m}\log\epsilon).
\end{align}
The coefficients $\mathcal{D}_{ab}^{(\textrm{r})}(r)$ and $\mathcal{D}_{ab}^{(\textrm{p})}(r)$ come from the expansion of the direct part of the parametrix $U/\sigma$ and correspond to the coefficients of terms in the expansion that are rational in $\Delta\tau^{2}$ and $s^{2}$ and those that are polynomial in $\Delta\tau^{2}$ and $s^{2}$, respectively. Similarly, $\mathcal{T}_{ab}^{(\textrm{r})}(r)$ and $\mathcal{T}_{ab}^{(\textrm{p})}(r)$ are the coefficients of terms that are rational and polynomial coming from the tail  $V\,\log(2\sigma/\ell^{2})$ of the parametrix, while  $\mathcal{T}_{ab}^{(\textrm{l})}(r)$ is the coefficient of terms in the tail that contain a logarithm. Expressions for the first few of these coefficients are given in Appendix~\ref{app:HCoeff}.

Formally speaking, this expansion need only be computed up to $m=1$ in order to remove the divergences from the Green function. Nevertheless, by expanding up to higher order terms, including terms that formally vanish in the coincidence limit, and expressing these terms as a mode-sum permits us to have precise control over the rate of convergence of the renormalized mode sum. On the other hand, if we only subtract the singular terms, the mode sum would indeed converge, however, it would converge slowly. Furthermore, the distinction between the different types of terms that occur in the expansion is an important one. In particular, the terms that are polynomial in the variables $\Delta\tau^{2}$ and $ s^{2}$ do not require a mode-sum representation. For example, a multipole expansion of terms that are polynomial in $s^{2}$ would involve finite multipole moments and hence cannot affect the large $l$ behaviour. Hence the terms above that are polynomial in the expansion parameters are left as-is. The rational and logarithmic terms will be expressed as mode sums. This is the focus of the next section. 

\subsection{Regularization Parameters for the Direct Part}
Starting with the direct part of the Hadamard parametrix, which involves terms of the form $\Delta\tau^{2a+2b}/s^{2+2b}$ for $a\ge b\ge 0$. We take as an ansatz for the mode-sum representation of this term the following
\begin{align}
\label{eq:directansatz}
    \frac{\Delta \tau^{2a+2b}}{s^{2+2b}}=\sum_{l=0}^{\infty}(2l+1)P_{l}(\cos\gamma)\int_{-\infty}^{\infty}d\omega e^{i\omega\Delta \tau}\Psi_{\omega l}(a,b|r),
\end{align}
where the generalized functions $\Psi_{\omega l}(a,b|r)$ we call the regularization parameters for the direct part. The goal is to invert the above ansatz and hence obtain an explicit closed-form representation for the regularization parameters. A double integral expression for the regularization parameters follows immediately from the completeness of the Legendre polynomials and the Fourier transform of the delta distribution, yielding
\begin{align}
  \Psi_{\omega l}(a,b|r)=\frac{1}{4\pi}\int_{-1}^{1}dx \int_{-\infty}^{\infty}d t \frac{e^{-i\omega t} t^{2a+2b} P_{l}(x)}{(f\,t^{2}+2r^{2}(1-x))^{1+b}}, 
\end{align}
where we have relabeled the integration variables by $x=\cos\gamma$ and $t=\Delta\tau$.
If we now define
\begin{align}
    \rho^{2}=\frac{f}{2r^{2}},
\end{align}
then we can rewrite our integral expression above as
\begin{align}
\label{eq:directint}
  \Psi_{\omega l}(a,b|r)=\frac{1}{4\pi}\frac{1}{(2r^{2})^{b+1}}\int_{-1}^{1}dx\,P_{l}(x)\frac{(-1)^{b}}{b!2^{b}}\nonumber\\
  \times\,\left(\frac{1}{\rho}\frac{\partial}{\partial\rho}\right)^{b}\int_{-\infty}^{\infty}d t \frac{e^{-i\omega t} t^{2a}}{(\rho^{2}\,t^{2}+1-x)}. 
\end{align}
Now it is this $t$-integral that distinguishes the regularization parameters in the Hartle-Hawking and the Boulware states. In the former, the $t$-integral above would be over a finite domain on account of the time periodicity of a thermal state. But in this case, the integral is over the entire real line which means that for $a\ge 1$, the integral does not converge in the sense of ordinary functions and must be considered in the sense of generalized functions. The most straightforward way to proceed is to start with the $a=0$ integral which is an ordinary function and is given by
\begin{align}
 \int_{-\infty}^{\infty}d t \frac{e^{-i\omega t} }{(\rho^{2}\,t^{2}+1-x)}=\pi\,\frac{e^{-|\omega|\sqrt{1-x}/\rho}}{\rho\sqrt{1-x}}.   
\end{align}
For $a\geq 1$, we can take $(2a)$-derivatives with respect to $\omega$ of the result above. Since the exponent on the right-hand side involves $|\omega|$ and not $\omega$, we pick up delta distributions and their derivatives. The result is
\begin{align}
 \int_{-\infty}^{\infty}d t \frac{e^{-i\omega t} t^{2a}}{(\rho^{2}\,t^{2}+1-x)}=\pi\,(-1)^{a}e^{-|\omega|\sqrt{1-x}/\rho}\frac{(1-x)^{a-1/2}}{\rho^{2a+1}}\nonumber\\
 -2\pi(-1)^{a}\sum_{p=1}^{a}\frac{(1-x)^{a-p}}{\rho^{2a-2p+2}}\delta^{(2p-2)}(\omega),   
\end{align}
which is clearly not an ordinary function. Substituting this back into (\ref{eq:directint}) gives
\begin{align}
     \Psi_{\omega l}(a,b|r)=\frac{1}{(2r^{2})^{b+1}}\frac{(-1)^{a+b}}{b!2^{b+2}}\left(\frac{1}{\rho}\frac{\partial}{\partial\rho}\right)^{b}\Big\{\frac{1}{\rho^{2a+1}}\mathcal{A}_{\omega l}(a|\rho)\nonumber\\
     -2\sum_{p=1}^{a}\frac{\delta^{(2p-2)}(\omega)}{\rho^{2a-2p+2}}\mathcal{B}_{l}(a,p)\Big\}, 
\end{align}
where
\begin{align}
    \mathcal{A}_{\omega l}(a|\rho)&=\int_{-1}^{1}P_{l}(x)e^{-|\omega|\sqrt{1-x}/\rho}(1-x)^{a-1/2}dx,\\
    \mathcal{B}_{ l}(a,p)&=\int_{-1}^{1}P_{l}(x)(1-x)^{a-p}dx.
\end{align}
The second integral here is a standard one. The result is that integral vanishes whenever $l\ge a-p+1$ and otherwise gives
\begin{align}
    \mathcal{B}_{l}(a,p)=\frac{(-1)^{l}2^{a-p+1}\Gamma(a-p+1)^{2}}{\Gamma(a-p+l+2)\Gamma(a-p+1-l)}.
\end{align}
The $\mathcal{A}_{\omega l}(a|\rho)$ integral can also be easily performed and there are several different equivalent representations. The tidiest form involves derivatives of the modified Bessel functions
\begin{align}
    \mathcal{A}_{\omega l}(a|\rho)=2^{a+3/2}\frac{\partial^{2a}}{\partial z^{2a}}I_{l+1/2}(z)K_{l+1/2}(z),
\end{align}
where 
\begin{align}
   z=\frac{|\omega|}{\sqrt{2}\,\rho}=\frac{|\omega|\,r}{\sqrt{f}}. 
\end{align}
\begin{widetext}
However, this is not the best representation for numerically computing this object. For numerical purposes, except when $z>>1$, we find the following representation much more efficient
\begin{align}
    \mathcal{A}_{\omega l}(a|\rho)=2^{a}\sqrt{2 \pi}(-1)^{l}\Bigg\{\Gamma(a+\tfrac{1}{2})^{2}{}_{2}\hat{F}_{3}\left(\begin{array}{c|}\{a+\tfrac{1}{2},a+\tfrac{1}{2}\}\\\{a+l+\tfrac{3}{2}, a-l+\tfrac{1}{2}, \tfrac{1}{2}\}\end{array}\,\,\frac{\omega^{2}}{2\rho^2}\right)\nonumber\\-\Gamma(a+1)^{2}\frac{|\omega|}{\sqrt{2}\,\rho}\,{}_{2}\hat{F}_{3}\left(\begin{array}{c|}\{a+1,a+1\}\\\{\tfrac{3}{2},a+l+2,a-l+1\}\end{array}\,\,\frac{\omega^{2}}{2\rho^2}\right)\Bigg\},
\end{align}
where ${}_{2}\hat{F}_{3}$ is the regularized general hypergeometric function \cite{NIST:DLMF}.
Now using the fact that,
\begin{align}
\label{eq:rhoderiv}
    \left(\frac{1}{\rho}\frac{\partial}{\partial\rho}\right)^{b}\rho^{-\alpha}=(-1)^{b}2^{b}\frac{\Gamma(\alpha/2+b)}{\Gamma(\alpha/2)}\rho^{-\alpha-2b},
\end{align}
we can easily express the derivatives in terms of hypergeometric functions. Putting all of the above together, and after some simplification, we obtain the following expression for the regularization parameters
\begin{align}
    \Psi_{\omega l}(a,b|r)=\mathcal{I}_{\omega l}(a,b|r)
    -\frac{(2r^{2})^{a}}{f^{a+b+1}}\frac{(-1)^{a+l}}{b!}\sum_{p=1}^{a}\frac{2^{a-p}(a-p)!(a-p+b)!}{(a-p+l+1)!(a-p-l)!}\left(\frac{f}{2r^{2}}\right)^{p}\delta^{(2p-2)}(\omega),
\end{align}
where, for later convenience, we have defined
\begin{align}
\label{eq:directregparam}
    \mathcal{I}_{\omega l}(a,b|r)=\frac{(2r^{2})^{a-1/2}}{f^{a+b+1/2}}\frac{2^{a-2}\sqrt{2 \pi}(-1)^{a+l}}{\Gamma(b+1)}\Bigg\{\Gamma(a+\tfrac{1}{2})\Gamma(a+b+\tfrac{1}{2}){}_{2}\hat{F}_{3}\left(\begin{array}{c|}\{a+\tfrac{1}{2},a+b+\tfrac{1}{2}\}\\\{a+l+\tfrac{3}{2}, a-l+\tfrac{1}{2}, \tfrac{1}{2}\}\end{array}\,\,\frac{\omega^{2}r^{2}}{f}\right)\nonumber\\
    -\Gamma(a+1)\Gamma(a+b+1)\frac{|\omega|r}{\sqrt{f}}\,{}_{2}\hat{F}_{3}\left(\begin{array}{c|}\{a+1,a+b+1\}\\\{\tfrac{3}{2},a+l+2,a-l+1\}\end{array}\,\,\frac{\omega^{2}r^{2}}{f}\right)\Bigg\}.
\end{align}
Substituting this back into (\ref{eq:directansatz}) and performing the $\omega$ integration in the terms involving delta distributions gives
\begin{align}
\label{eq:directmodesum}
    \frac{\Delta\tau^{2a+2b}}{s^{2+2b}}=\sum_{l=0}^{\infty}(2l+1)P_{l}(\cos\gamma)\int_{-\infty}^{\infty}d\omega \,e^{i\omega\Delta\tau}\mathcal{I}_{\omega l}(a,b|r)
    +\frac{1}{f^{b+1}}\sum_{p=1}^{a}\frac{(a-p+b)!}{b!(a-p)!}\left(\frac{2 r^{2}}{f}\right)^{a-p}\hspace{-2mm}\Delta\tau^{2p-2}(\cos\gamma-1)^{a-p}.
\end{align}
\end{widetext}
In arriving at the last line of this expression, we made use of the identity
    \begin{align}
\label{eq:legendreidentity}
    \sum_{l=0}^{n}\frac{(-1)^{l}(2l+1)P_{l}(\cos\gamma)}{(n+l+1)!(n-l)!}=\frac{(1-\cos\gamma)^{n}}{2^{n}(n!)^{2}},
\end{align}
which can be proved by induction. In the coincidence limit, the last line of (\ref{eq:directmodesum}) is nonzero only when $p=a=1$.

\subsection{Regularization Parameters for the Logarithmic Terms in the Tail}
In this subsection, we derive a mode-sum representation for the terms in the tail part of the Hadamard parametrix which in the partial coincidence limit $\Delta r=0$ involves terms of the form $\Delta \tau^{2b}s^{2a-2b}\log(s^{2}/\ell^{2})$. For later convenience, we will write the logarithmic terms as
\begin{align}
\label{eq:tailmodesumansatz}
    \Delta \tau^{2b}s^{2a-2b}\log(s^{2}/\ell^{2})=\Delta \tau^{2b}s^{2a-2b}\log(\rho^{2}\Delta \tau^{2}+1-x)\nonumber\\
    +\Delta \tau^{2b}s^{2a-2b}\log(2r^{2}/\ell^{2}).
\end{align}
The second term here is polynomial in $\Delta\tau^{2}$ and $s^{2}$ and requires no mode-sum representation. This term contains the arbitrary lengthscale $\ell$ which encodes part of the renormalization ambiguity (see, for example, Refs.~\cite{HollandsWald2005, WaldBook:1994, DecaniniFolacci2008}). This term does not contribute to the vacuum polarization since $a\ge 1$ in these terms and hence they all vanish in the coincidence limit. For the first term above, we assume the mode-sum ansatz
\begin{align}
    &\Delta\tau^{2b}s^{2a-2b}\log(\rho^{2}\Delta\tau^{2}+1-\cos\gamma)\nonumber\\
   & =\sum_{l=0}^{\infty}(2l+1)P_{l}(\cos\gamma)\int_{-\infty}^{\infty}d\omega\,e^{i\omega\,\Delta\tau}\,\chi_{\omega l}(a,b\,|r)
\end{align}
and as before, we can invert to obtain a formal double integral expression for the regularization parameters
\begin{align}
\chi_{\omega l}&(a,b|r)=\frac{(2 r^{2})^{a-b}}{4\pi}\int_{-1}^{1}dx P_{l}(x)\nonumber\\
&\times\lim_{\alpha\to a}\frac{\partial}{\partial \alpha}\int_{-\infty}^{\infty}e^{-i\omega t}t^{2b}(\rho^{2}t^{2}+1-x)^{\alpha-b}dt,
\end{align}
where we have expressed the log terms as the limit of a derivative with respect to an exponent $\alpha$, and where $t$ and $x$ are defined as before. The time integral above is a complicated generalized function but we can express it in terms of the derivative of a simpler integral as
\begin{align}
    &\chi_{\omega l}(a,b\,|r)=\frac{(2 r^{2})^{a-b}}{4\pi}\frac{1}{2^{b}}\left(\frac{1}{\rho}\frac{\partial}{\partial \rho}\right)^{b}\int_{-1}^{1}dx P_{l}(x)\nonumber\\
&\times\lim_{\alpha\to a}\frac{\partial}{\partial \alpha}\frac{\Gamma(\alpha-b+1)}{\Gamma(\alpha+1)}\int_{-\infty}^{\infty}e^{-i\omega t}(\rho^{2}t^{2}+1-x)^{\alpha}dt.
\end{align}
The advantage of this form is that the time integral can now be performed explicitly yielding,
\begin{align}
\label{eq:log_expr1}
    \chi_{\omega l}&(a,b\,|r)=\frac{(2r^{2})^{a-b}}{\sqrt{2\pi}}\frac{1}{2^{b}}\left(\frac{1}{\rho}\frac{\partial}{\partial \rho}\right)^{b}\nonumber\\
    &\times \int_{-1}^{1}dx P_{l}(x)\mathcal{F}_{\omega l}(a,b\,|x,r)
\end{align}
where
\begin{widetext}
\begin{align}
   \mathcal{F}_{\omega l}(a,b\,|x,r)=\lim_{\alpha\to a} \frac{\partial}{\partial \alpha}\left\{\frac{2^{\alpha}\rho^{\alpha-1/2}\Gamma(\alpha-b+1)}{|\omega|^{\alpha+1/2}\Gamma(-\alpha)\Gamma(\alpha+1)}(1-x)^{\alpha/2+1/4}K_{\alpha+1/2}\left(\frac{|\omega|\sqrt{1-x}}{\rho}\right)\right\},
\end{align}
and $K_{\alpha+1/2}(z)$ is the modified Bessel function of the second kind. From the series representation of the Bessel function, we obtain
\begin{align}
\label{eq:Fseries}
    \mathcal{F}_{\omega l}(a,b\,|x,r)=\frac{\pi}{\sqrt{2}}\lim_{\alpha\to a}\frac{\partial}{\partial \alpha}\Bigg\{\frac{\Gamma(\alpha-b+1)}{\Gamma(\alpha+1)}\frac{\sec\pi\alpha\, (2\rho)^{2\alpha}}{\Gamma(-\alpha)|\omega|^{2\alpha+1}}\sum_{k=0}^{\infty}\frac{1}{k!\Gamma(-\alpha+k+\tfrac{1}{2})}\frac{\omega^{2k}(1-x)^{k}}{(2\rho)^{2k}}\nonumber\\
   -\frac{\Gamma(\alpha-b+1)}{\Gamma(\alpha+1)}\frac{\sec\pi\alpha\, (1-x)^{\alpha+1/2}}{\Gamma(-\alpha)2\rho}\sum_{k=0}^{\infty}\frac{1}{k!\Gamma(\alpha+k+\tfrac{3}{2})}\frac{\omega^{2k}(1-x)^{k}}{(2\rho)^{2k}} \Bigg\}.
\end{align}
\end{widetext}
The second term here is straightforward to deal with in the limit as $\alpha\to a$. In particular, it is evidently an entire function of $\alpha$. In fact, if we consider the Taylor series for $1/\Gamma(-\alpha)$ about a positive integer $a$, we have 
\begin{align}
\label{eq:gammapole}
    \frac{1}{\Gamma(-\alpha)}=&(-1)^{a+1}a!(\alpha-a)+(-1)^{a+1}a!\psi(a+1)(\alpha-a)^{2}\nonumber\\
    &+\mathcal{O}(\alpha-a)^{3},
\end{align}
where $\psi(z)=\Gamma'(z)/\Gamma(z)$ is the polygamma function, and so it is clear that the only term in the last line of (\ref{eq:Fseries}) that survives the limit $\alpha\to a$ is when the $\alpha$-derivative acts on $1/\Gamma(-\alpha)$.

The first term in (\ref{eq:Fseries}) is more subtle owing to the presence of  the  $|\omega|^{-2\alpha-1}$ term. Clearly this is not an ordinary function since it is singular at $\omega=0$. It is a generalized function whose definition involves a regularization at $\omega=0$ (not to be confused with the renormalization prescription required to regularize the coincidence limit of the point-split two-point function). One way to treat this is to simply introduce an infra-red cutoff in the $\omega$-integral in (\ref{eq:tailmodesumansatz}), but this is not the rigorous approach. We will see that, if we treat $|\omega|^{-2\alpha-1}$ rigorously as a generalized function, a cutoff will emerge naturally and indeed it will be clear that the overall result will be independent of this choice. The salient point is that, as a function of the exponent $\alpha$, the definition (which involves its regularization as a distribution) of $|\omega|^{-2\alpha-1}$ has a simple pole at integer values of $\alpha$. However, this pole is canceled by the zero in $1/\Gamma(-\alpha)$ at integer values (see Eq.~(\ref{eq:gammapole})) so that the combination $1/(\Gamma(-\alpha)|\omega|^{2\alpha+1})$ is actually an entire function. We require an expansion of this combination about the integer $\alpha=a$ up to order $(\alpha-a)^{2}$ if we are to compute the limit appearing in (\ref{eq:Fseries}). We already have a series for $1/\Gamma(-\alpha)$. Next, we note the Laurent series for $|\omega|^{-2\alpha-1}$ which can be found in Ref.~\cite{GelfandShilovVol1},
\begin{align}
\label{eq:omegaLaurent}
    |\omega|^{-2\alpha-1}&=-\frac{\delta^{(2a)}(\omega)}{(2a)!(\alpha-a)}+|\omega|^{-2a-1}+\frac{\delta^{(2a)}(\omega)}{(2a)!}2\log\lambda\nonumber\\
    &+\mathcal{O}(\alpha-a),
\end{align}
where the $\lambda$ appearing here is an arbitrary inverse lengthscale required because $\omega$ has dimensions of inverse length. This inverse lengthscale will be shown later to be related to an infra-red cutoff in certain integrals over the frequency. The corresponding result for Eq.~(\ref{eq:omegaLaurent}) in Ref.~\cite{GelfandShilovVol1} is actually for a dimensionless generalized function and so doesn't contain the additional $\log\lambda$ term, but applying the result from \cite{GelfandShilovVol1} to $\hat{\omega}=\omega/\lambda$ gives Eq.~(\ref{eq:omegaLaurent}). It is essential to understand that $|\omega|^{-2a-1}$ appearing on the right-hand side is to be understood in terms of its regularization as a generalized function. For example, for $\varphi(\omega)$ a test function of compact support, we have 
\begin{align}
\label{eq:regnegativepowers}
    (|\omega|^{-2a-1},\varphi(\omega))=\int_{0}^{\infty}d\omega\,\omega^{-2a-1}\Big[\varphi(\omega)+\varphi(-\omega)\nonumber\\
    -2\sum_{m=0}^{a-1}\frac{\varphi^{(2m)}(0)}{(2m)!}\omega^{2m}-2\frac{\varphi^{(2a)}(0)}{(2a)!}\omega^{2a}\theta(\lambda-\omega)\Big],
\end{align}
where $\theta(z)$ is the step function. The last term in the regularization requires this step function since otherwise the integral would not converge at infinity.

Now combining (\ref{eq:gammapole}) and (\ref{eq:omegaLaurent}) gives us the desired expansion
\begin{align}
  &  \frac{1}{\Gamma(-\alpha)|\omega|^{2\alpha+1}}=\frac{(-1)^{a}a!}{(2a)!}\delta^{(2a)}(\omega)\nonumber\\
   & +(\alpha-a)(-1)^{a}a!\Big[\frac{(\psi(a+1)-2\log\lambda)}{(2a)!}\delta^{(2a)}(\omega)-\frac{1}{|\omega|^{2a+1}}\Big]\nonumber\\
   & +\mathcal{O}(\alpha-a)^{2}.
\end{align}
Employing these limits in Eq.~(\ref{eq:Fseries}) gives, after some tedious simplifications,
\begin{widetext}
\begin{align}
    \mathcal{F}_{\omega l}(a,b\,|x,r)=\frac{\pi}{\sqrt{2}}(a-b)!\Bigg\{((\psi(a-b+1)+\psi(\tfrac{1}{2}-a+k)+2\ln(2\rho/\lambda))\sum_{k=0}^{a}\frac{\delta^{(2a-2k)}(\omega)(1-x)^{k}(2\rho)^{2a-2k}}{k!(2a-2k)!\Gamma(k-a+\tfrac{1}{2})}\nonumber\\
    -\sum_{k=0}^{\infty}\frac{|\omega|^{2k-2a-1}(1-x)^{k}(2\rho)^{2a-2k}}{k!\Gamma(k-a+\tfrac{1}{2})}+\sum_{k=0}^{\infty}\frac{\omega^{2k}(1-x)^{k+a+1/2}}{k!\Gamma(k+a+\tfrac{3}{2})(2\rho)^{2k+1}}\Bigg\},
\end{align}
where we also made use of the distributional identity
\begin{align}
   \omega^{2k} \delta^{(2a)}(\omega)=\begin{cases} \frac{(2a)!}{(2a-2k)!}\delta^{(2a-2k)}(\omega)\qquad & k\le a\\ 0 &k>a \end{cases}
\end{align}
to truncate the infinite sums involving delta distributions. Substituting this expression back into (\ref{eq:log_expr1}) and performing the integration yields
\begin{align}
\label{eq:chitemp}
    \chi_{\omega l}(a,b\,|r)&=\sqrt{\pi}(2r^{2})^{a-b}(a-b)!\frac{(-1)^{l}}{2^{b}}\left(\frac{1}{\rho}\frac{\partial}{\partial\rho}\right)^{b}\Bigg\{\sum_{k=0}^{\infty}\frac{\Gamma(k+a+\tfrac{3}{2})2^{k+a+3/2}\omega^{2k}}{k!\,\Gamma(k+a+l+\tfrac{5}{2})\,\Gamma(k+a-l+\tfrac{3}{2})(2\rho)^{2k+1}}\nonumber\\
    &-\sum_{k=0}^{\infty}\frac{k!2^{k}(2\rho)^{2a-2k}|\omega|^{2k-2a-1}}{(k+l+1)!(k-l)!\Gamma(k-a+\tfrac{1}{2})}\nonumber\\
    &+[\psi(a-b+1)+\psi(k-a+\tfrac{1}{2})+2\log(2\rho/\lambda)]\sum_{k=0}^{a}\frac{k!2^{k}(2\rho)^{2a-2k}\delta^{(2a-2k)}(\omega)}{(2a-2k)!(k+l+1)!(k-l)!\Gamma(k-a+\tfrac{1}{2})}\Bigg\}.
\end{align}
The second term is a generalized function of $\omega$  whenever $k\le a$ and an ordinary function otherwise. In the former case, it is understood in the sense of (\ref{eq:regnegativepowers}). It would be better if we could could express the definition in (\ref{eq:regnegativepowers}) in a way that is independent of the test function. It can be shown that the regularization in Eq.~(\ref{eq:regnegativepowers}) is equivalent to defining the generalized function as
\begin{align}
\label{eq:regnegpowerscutoff}
    \frac{1}{|\omega|^{2j+1}}=\frac{\theta(\omega^{2}-\lambda^{2})}{|\omega|^{2j+1}}+2\sum_{k\ne j}^{\infty}\frac{\lambda^{2k-2j}\delta^{(2k)}(\omega)}{(2k)!(2k-2j)},\qquad j>0\in\mathbb{Z}
\end{align}
where $\lambda>0$ is the arbitrary cut-off. This can be straightforwardly proven by splitting the integral in (\ref{eq:regnegativepowers}) into intervals  $(0,\lambda)$ and $(\lambda,\infty)$, and then employing a Taylor series about $\omega=0$ for the test function on the interval $(0,\lambda)$. Employing definition (\ref{eq:regnegpowerscutoff}) in the second term of (\ref{eq:chitemp}) for $k\le a$ gives
\begin{align}
\label{eq:chitemp2}
    \chi_{\omega l}(a,b\,|r)&=\sqrt{\pi}(2r^{2})^{a-b}(a-b)!\frac{(-1)^{l}}{2^{b}}\left(\frac{1}{\rho}\frac{\partial}{\partial\rho}\right)^{b}\Bigg\{\sum_{k=0}^{\infty}\frac{\Gamma(k+a+\tfrac{3}{2})2^{k+a+3/2}\omega^{2k}}{k!\,\Gamma(k+a+l+\tfrac{5}{2})\,\Gamma(k+a-l+\tfrac{3}{2})(2\rho)^{2k+1}}\nonumber\\
    &-\theta(\omega^{2}-\lambda^{2})\sum_{k=0}^{a}\frac{k!2^{k}(2\rho)^{2a-2k}|\omega|^{2k-2a-1}}{(k+l+1)!(k-l)!\Gamma(k-a+\tfrac{1}{2})}-\sum_{k=0}^{\infty}\frac{(k+a+1)!2^{k+a+1}|\omega|^{2k+1}}{(k+a+l+2)!(k+a+1-l)!\Gamma(k+\tfrac{3}{2})(2\rho)^{2k+2}}\nonumber\\
    &-2\sum_{k=0}^{a}\frac{k!2^{k}(2\rho)^{2a-2k}}{(k+l+1)!(k-l)!\Gamma(k-a+\tfrac{1}{2})}\sum_{p\ne a-k}^{\infty}\frac{\lambda^{2p+2k-2a}\delta^{(2p)}(\omega)}{(2p)!(2p+2k-2a)}\nonumber\\
    &+[\psi(a-b+1)+\psi(k-a+\tfrac{1}{2})+2\log(2\rho/\lambda)]\sum_{k=0}^{a}\frac{k!2^{k}(2\rho)^{2a-2k}\delta^{(2a-2k)}(\omega)}{(2a-2k)!(k+l+1)!(k-l)!\Gamma(k-a+\tfrac{1}{2})}\Bigg\}.
\end{align}
Finally computing the derivatives and simplifying gives
\begin{align}
    \chi_{\omega l}(a,b\,|r)&=\mathcal{J}_{\omega l}(a,b\,|r)+\sqrt{\pi}(2r^{2}\rho^{2})^{a-b}(a-b)!\,2^{2a}(-1)^{l}\nonumber\\
   & \times\Bigg\{\sum_{k=0}^{a}\frac{k!(a-k)!\delta^{(2a-2k)}(\omega)\left[\psi(a-b+1) -\psi(a-b-k+1)+\psi(a-k+1)+\psi(k-a+\tfrac{1}{2})+2\log(\tfrac{2\rho}{\lambda})\right]}{(a-k-b)!(2a-2k)!\Gamma(k-a+\tfrac{1}{2})(k+l+1)!(k-l)!(2\rho^{2})^{k}}\nonumber\\
   &-2\sum_{k=0}^{a}\frac{(a-k)!k!}{(a-k-b)!\Gamma(k-a+\tfrac{1}{2})(k+l+1)!(k-l)!(2\rho^{2})^{k}} \sum_{\substack{p=0\\p\ne a-k}}^{\infty}\frac{\delta^{(2p)}(\omega)\lambda^{2p-2a+2k}}{(2p)!(2p-2a+2k)}
    \Bigg\},
\end{align}
where
\begin{align}
    \mathcal{J}_{\omega l}(a,b\,|r)&=\sqrt{\pi}(2r^{2}\rho^{2})^{a-b}(a-b)!2^{2a}(-1)^{l}\Bigg\{-\sum_{k=0}^{a-b}\frac{\theta\left(\omega^{2}-\lambda^{2}\right)(a-k)!k!|\omega|^{2k-2a-1}}{(a-k-b)!\Gamma(k-a+\tfrac{1}{2})(k+l+1)!(k-l)!(2\rho^{2})^{k}}\nonumber\\
    &-\frac{(-1)^{b}}{(2\rho^{2})^{a+1}}|\omega|(a+1)!b!\,{}_{2}\hat{F}_{3}\left(\begin{array}{c|}\{a+2,b+1\}\\\{\tfrac{3}{2},a-l+2,a+l+3\}\end{array}\,\,\frac{\omega^{2}}{2\rho^{2}}\right)\nonumber\\
    &+\frac{(-1)^{b}}{2^{a+1/2}\rho^{2a+1}}\Gamma(a+\tfrac{3}{2})\Gamma(b+\tfrac{1}{2}){}_{2}\hat{F}_{3}\left(\begin{array}{c|}\{a+\tfrac{3}{2},b+\tfrac{1}{2}\}\\\{\tfrac{1}{2},a-l+\tfrac{3}{2},a+l+\tfrac{5}{2}\}\end{array}\,\,\frac{\omega^{2}}{2\rho^{2}}\right)\Bigg\}.
\end{align}
Putting this together gives
\begin{align}
\label{eq:logterms}
\Delta\tau^{2b}s^{2a-2b}&\log(\rho^{2}\Delta\tau+1-\cos\gamma)=\int_{-\infty}^{\infty}d\omega\,e^{i\omega\Delta\tau}\sum_{l=0}^{\infty}(2l+1)P_{l}(\cos\gamma)\chi_{\omega l}(r)\nonumber\\
&=\int_{-\infty}^{\infty}d\omega\,e^{i\omega\Delta\tau}\sum_{l=0}^{\infty}(2l+1)P_{l}(\cos\gamma)\mathcal{J}_{\omega l}(a,b\,|r)\nonumber\\
&+\sqrt{\pi}(2r^{2}\rho^{2})^{a-b}(a-b)! 2^{2a}\Bigg\{\sum_{k=0}^{a-b}\frac{(-1)^{a-k}(a-k)!\Delta\tau^{2a-2k}(1-\cos\gamma)^{k}}{2^{k}k!(a-k-b)!(2a-2k)!\Gamma(k-a+\tfrac{1}{2})(2\rho^{2})^{k}}\Big[[\psi(a-b+1)\nonumber\\
&-\psi(a-b-k+1)+\psi(a-k+1)+\psi(k-a+\tfrac{1}{2})+2\log(2\rho/\lambda)\Big]\nonumber\\
&-\sum_{k=1}^{b}\frac{(-1)^{b}(b-k)!(k-1)!\Delta\tau^{2b-2k}(1-\cos\gamma)^{k+a-b}}{2^{k+a-b}(k+a-b)!(2b-2k)!\Gamma(k-b+\tfrac{1}{2})(2\rho^{2})^{k+a-b}}\nonumber\\
&-2\sum_{k=0}^{a-b}\frac{(a-k)!(1-\cos\gamma)^{k}}{2^{k}k!(a-k-b)!\Gamma(k-a+\tfrac{1}{2})(2\rho^{2})^{k}}\sum_{\substack{p=0\\p\ne a-k}}^{m-k}\frac{(-1)^{p}\Delta\tau^{2p}\lambda^{2p-2a+2k}}{(2p)!(2p-2a+2k)}\Bigg\}.
\end{align}
\end{widetext}
Even though the sum over $p$ in equation \eqref{eq:logterms} is technically up to $\infty$, we actually only need terms up to order $\epsilon^{2m}$ in the Hadamard parametrix. In that expression, the order of the expansion is given by $(1-\cos\gamma)^k\Delta\tau^{2p}$ and is $\epsilon^{2k+2p}$. Hence, we have truncated the sum at $p=m-k$. In arriving at the expression above, we have also made use of the identity \eqref{eq:legendreidentity}.

\subsection{Mode-sum expression for the parametrix}
We now have mode-sum representations of all the singular terms in \eqref{eq:hadamard_mode_sum_1} yielding the following mode-sum representation of the Hadamard parametrix
\begin{widetext}
\begin{align}
\label{eq:hadamard_mode_sum_2}
    K(x,x')&=\frac{1}{8\pi^{2}}\int_{-\infty}^{\infty}d\omega\,e^{i\omega\Delta\tau}\sum_{l=0}^{\infty}(2l+1)P_{l}(\cos\gamma)k_{\omega l}^{(m)}(r)+\frac{1}{8\pi^{2}}\Bigg\{\sum_{a=1}^{m}\sum_{b=1}^{a}\mathcal{D}_{ab}^{(\textrm{p})}(r)\Delta\tau^{2a-2b}s^{2b-2}\nonumber\\
    &+\sum_{a=1}^{m-1}\sum_{b=0}^{a}\mathcal{T}_{ab}^{(\textrm{p})}(r)\Delta\tau^{2b}s^{2a-2b}+\sum_{a=0}^{m-1}\sum_{b=0}^{a}\mathcal{T}_{ab}^{(\textrm{l})}(r)\Delta\tau^{2b}s^{2a-2b}\log(4f/(\ell^{2}\lambda^2))\nonumber\\
    &+\sum_{a=1}^{m}\sum_{b=0}^{a}\frac{\mathcal{D}_{ab}^{(\textrm{r})}(r)\,(2r^{2})^{-b-1}}{b!\,(\rho^{2})^{a+b+1}}\sum_{p=1}^{a}\frac{(-1)^{a+p}(a-p+b)!\rho^{2p}}{(a-p)!}\Delta\tau^{2p-2}(1-\cos\gamma)^{a-p}\nonumber\\
    &+\sum_{a=1}^{m-1}\sum_{b=0}^{a-1}\frac{\mathcal{T}_{ab}^{(\textrm{r})}(r)\,(2r^{2})^{-b-1}}{b!\,(\rho^{2})^{a+b+2}}\sum_{p=1}^{a+1}\frac{(-1)^{a+1+p}(a+1-p+b)!\rho^{2p}}{(a+1-p)!}\Delta\tau^{2p-2}(1-\cos\gamma)^{a+1-p}\nonumber\\
    &+\sum_{a=0}^{m-1}\sum_{b=0}^{a}\mathcal{T}_{ab}^{(\textrm{l})}(r)\sqrt{\pi}(2r^{2}\rho^{2})^{a-b}(a-b)! 2^{2a}\Bigg[\sum_{k=0}^{a-b}\frac{(-1)^{a-k}(a-k)!\Delta\tau^{2a-2k}(1-\cos\gamma)^{k}}{2^{k}k!(a-k-b)!(2a-2k)!\Gamma(k-a+\tfrac{1}{2})(2\rho^{2})^{k}}\Big[\psi(a-b+1)\nonumber\\
    &-\psi(a-b-k+1)+\psi(a-k+1)+\psi(k-a+\tfrac{1}{2})\Big]\nonumber\\
    &-\sum_{k=1}^{b}\frac{(-1)^{b}(b-k)!(k-1)!\Delta\tau^{2b-2k}(1-\cos\gamma)^{k+a-b}}{2^{k+a-b}(k+a-b)!(2b-2k)!\Gamma(k-b+\tfrac{1}{2})(2\rho^{2})^{k+a-b}}\nonumber\\
    &-2\sum_{k=0}^{a-b}\frac{(a-k)!(1-\cos\gamma)^{k}}{2^{k}k!(a-k-b)!\Gamma(k-a+\tfrac{1}{2})(2\rho^{2})^{k}}\sum_{\substack{p=0\\p\ne a-k}}^{m-k}\frac{(-1)^{p}\Delta\tau^{2p}\lambda^{2p-2a+2k}}{(2p)!(2p-2a+2k)}\Bigg]\Bigg\}+\mathcal{O}(\epsilon^{2m}\log\epsilon),
\end{align}
\end{widetext}
where we have defined for compactness
\begin{align}
\label{eq:k_oml_m}
    k_{\omega l}^{(m)}(r)&=\sum_{a=0}^{m}\sum_{b=0}^{a}\mathcal{D}_{ab}^{(\textrm{r})}(r)\mathcal{I}_{\omega l}(a,b|r)\nonumber\\
    &+\sum_{a=1}^{m-1}\sum_{b=0}^{a-1}\mathcal{T}_{ab}^{(\textrm{r})}(r)\mathcal{I}_{\omega l}(a+1,b|r)\nonumber\\
    &+\sum_{a=0}^{m-1}\sum_{b=0}^{a}\mathcal{T}_{ab}^{(\textrm{l})}(r)\mathcal{J}_{\omega l}(a,b\,|r).
\end{align}
While the expression \eqref{eq:hadamard_mode_sum_2} looks very complicated, we note that many of the terms vanish in the coincidence limit.

\section{Vacuum polarization and components of the RSET}
\label{sec:Reg}
When computing vacuum polarization, the coincidence limit kills almost all of the finite terms in the Hadamard parametrix above, we obtain
\begin{align}
\label{eq:vacpolren}
    \langle \hat{\phi}^{2}\rangle_{\textrm{ren}}=&\frac{1}{8\pi^{2}}\Bigg\{\int_{-\infty}^{\infty}\sum_{l=0}^{\infty}(2l+1)\left(g_{\omega l}(r)-k_{\omega l}^{(m)}(r)\right)d\omega\nonumber\\
    &-\sqrt{\pi}\sum_{a=1}^{m-1}\sum_{b=0}^{a}\mathcal{T}_{ab}^{(\textrm{l})}(r)\,\frac{(a-1)!2^{2a}f^{a-b}}{\Gamma(-a+\tfrac{1}{2})\lambda^{2a}}\Bigg\}\nonumber\\
     &-\frac{1}{8\pi^{2}}\Bigg[\mathcal{D}_{11}^{(\textrm{p})}(r)+\frac{\mathcal{D}_{10}^{(\textrm{r})}(r)}{f}+\frac{\mathcal{D}_{11}^{(\textrm{r})}(r)}{f^2}\nonumber\\
     &+\mathcal{T}_{00}^{(\textrm{l})}(r)(\log( f /(\ell^{2}\lambda^2))-2\gamma_{\textrm{E}})\Bigg].
\end{align}
The mode-sum terms above can be made to converge as fast as we like by choosing $m$ appropriately large. It can be shown that the renormalized vacuum polarization is completely independent on the choice of the cutoff (we report the calculations in appendix \ref{app:cutoff} for clarity). This property can also be proven for the regularization terms for the stress energy tensor.

The components of the RSET may be written in the form \cite{taylorbreenottewill:2021}:
\begin{align}
	\label{eq:RSET}
	\langle \hat{T}^{\mu}{}_{\nu}\rangle_{\textrm{ren}} & = - \mathsf{w}^{\mu}{}_{\nu}   -   (\xi-\tfrac{1}{2})\mathsf{w}^{;\mu}{}_{;\nu}+ (\xi-\tfrac{1}{4}) \square \mathsf{w} \,\delta^{\mu}{}_{\nu}  \nonumber\\
	&\qquad +\xi R^{\mu}{}_{\nu} \mathsf{w} -  \frac{1}{8\pi^2} v_1 \delta^{\mu}{}_{\nu}.
\end{align}
where:
\begin{align}
        &\mathsf{w}(r) =\langle\hat\phi^2\rangle_{\textrm{ren}}(r),\nonumber\\
	&\mathsf{w}_{\mu\nu}(x)\equiv\lim_{x \to x'}\left[W(x,x')_{;\mu\nu}\right],\nonumber\\
	&W(x,x')=G(x,x')-K(x,x'),
\end{align}
and
\begin{align}
	v_1 &= \tfrac{1}{720}R_{\mu\nu\rho\sigma}R^{\mu\nu\rho\sigma}- \tfrac{1}{720}R_{\mu\nu}R^{\mu\nu}- \tfrac{1}{24}(\xi-\tfrac{1}{5})\square R\nonumber\\
	&\quad  + \tfrac{1}{8}(\xi-\tfrac{1}{6})^2R^2 + \tfrac{1}{4}\mu^2(\xi-\tfrac{1}{6}) R +\tfrac{1}{8}\mu^4,
\end{align}
which must be included in $(\ref{eq:RSET})$ to ensure the RSET is conserved \cite{WaldBook:1994}.

Due to the symmetry of the background spacetime $\mathsf{w}$ is a function of $r$ only, therefore once it has been calculated numerically to high accuracy on a suitably dense grid, the derivatives of $\mathsf{w}$ appearing in Eq.~\eqref{eq:RSET} are now easily and accurately obtained by differentiating an interpolation function for $\mathsf{w}$. Considering the components of $\mathsf{w}^{a}{}_{b}$, we have, by virtue of the wave equation satisified by $W(x,x')$, that~\cite{BrownOttewill:1986}: 
\begin{align}
	\mathsf{w}^{r}{}_{r}  = -\mathsf{w}^{\tau}{}_{\tau}-\mathsf{w}^{\theta}{}_{\theta}-\mathsf{w}^{\phi}{}_{\phi} +  \xi R \mathsf{w}  +\mu^2 \mathsf{w}  -\frac{3}{4\pi^2}v_1.
\end{align}
For the remaining non-zero components, $\mathsf{w}^{\tau}{}_{\tau}$ and \mbox{$\mathsf{w}^{\theta}{}_{\theta}=\mathsf{w}^{\phi}{}_{\phi}$}, it is advantageous to express these in terms of mixed derivatives at $x$ and $x'$ of $W(x,x')$ and derivatives of $\mathsf{w}(r) =\langle\hat\phi^2\rangle_{\textrm{ren}}$, using Synge's Rule \cite{taylorbreenottewill:2021}:
\begin{align}
\lim_{x \to x'}	\left[W(x,x')_{;\mu'\nu}\right] =  \tfrac{1}{2} \mathsf{w}_{;\mu\nu}(x) - \mathsf{w}_{\mu\nu}(x).
\end{align}  
The required mixed time derivatives and mixed angular derivatives may be obtained by taking appropriate derivatives of Eq (\ref{eq:hadamard_mode_sum_1}). Therefore, the RSET is determined by the calculation of three mode sums Eq.(\ref{eq:vacpolren}) along with the following two mode sums:
\begin{align}\label{eq:gttren}
	[g^{\tau \tau'}&W_{,\tau\tau'}]=\nonumber\\&\frac{1}{8\pi^{2} f}\Bigg\{\int_{-\infty}^{\infty}\sum_{l=0}^{\infty}(2l+1)\omega^2\left(g_{\omega l}-k_{\omega l}^{(m)}\right)d\omega\nonumber\\
&-\sqrt{\pi}\sum_{\substack{a=0\\a\ne 1}}^{m-1}\sum_{b=0}^{a}\mathcal{T}_{ab}^{(\textrm{l})}(r)\,\frac{a!2^{2a}f^{a-b}}{\Gamma(\tfrac{1}{2}-a)(a-1)\lambda^{2a-2}}\Bigg\}\nonumber\\&
	+\frac{1}{4\pi^{2}}\Bigg[\mathcal{T}_{10}^{(\textrm{p})}+\mathcal{D}_{22}^{(\textrm{p})}+\frac{1}{f}(\mathcal{T}_{11}^{(\textrm{p})}+\mathcal{D}_{21}^{(\textrm{p})}) \nonumber\\&+\frac{\mathcal{D}_{20}^{(\textrm{r})}}{f^2}+\frac{\mathcal{D}_{21}^{(\textrm{r})}}{f^3}+\frac{\mathcal{D}_{22}^{(\textrm{r})}}{f^4}+\frac{\mathcal{T}_{10}^{(\textrm{r})}}{f^2} \nonumber\\&+(f\mathcal{T}_{10}^{(\textrm{l})}+\mathcal{T}_{11}^{(\textrm{l})})(\log( f /(\ell^{2}\lambda^{2}))-2\gamma_{\textrm{E}}+3)\Bigg],
\end{align}
\begin{align}\label{eq:gphiphiren}
	[g^{\phi \phi'}&W_{,\phi\phi'}]=\nonumber\\&\frac{1}{16\pi^{2}r^{2}}\Bigg\{\int_{-\infty}^{\infty}\sum_{l=0}^{\infty}(2l+1)l(l+1)\left(g_{\omega l}-k_{\omega l}^{(m)}\right)d\omega\nonumber\\&+\sqrt{\pi}\sum_{a=2}^{m-1}\sum_{b=0}^{a}\mathcal{T}_{ab}^{(\textrm{l})}\,\frac{r^2(a-b)(a-2)!2^{2a}f^{a-b-1}}{\Gamma(\tfrac{3}{2}-a)\lambda^{2a-2}}\Bigg\}\nonumber\\&
    +\frac{1}{4\pi^{2}}\Bigg[\mathcal{D}_{22}^{(\textrm{p})}+\mathcal{T}_{10}^{(\textrm{p})}-\frac{\mathcal{D}_{20}^{(\textrm{r})}}{f^2}-2\frac{\mathcal{D}_{21}^{(\textrm{r})}}{f^3}-3\frac{\mathcal{D}_{22}^{(\textrm{r})}}{f^4}\nonumber\\&-\frac{\mathcal{T}_{10}^{(\textrm{r})}}{f^2} +\frac{\mathcal{T}_{11}^{(\textrm{l})}}{f} +r^2\mathcal{T}_{10}^{(\textrm{l})}(\log( f /(\ell^{2}\lambda^{2}))-2\gamma_{\textrm{E}}+1)\Bigg].
\end{align}
Inserting the expressions in Eqs. (\ref{eq:vacpolren}),  (\ref{eq:gttren}),  
 (\ref{eq:gphiphiren}) into Eq.~(\ref{eq:RSET}) results in a natural splitting of the RSET components into a numeric and an analytic part:
\begin{align}
\label{eq:Tsplit}
	\langle \hat{T}^{\mu}{}_{\nu}\rangle_{\textrm{ren}}=	\langle \hat{T}^{\mu}{}_{\nu}\rangle_{\textrm{numeric}}+ 	\langle \hat{T}^{\mu}{}_{\nu}\rangle_{\textrm{analytic}},
\end{align}
 where $\langle \hat{T}^{\mu}{}_{\nu}\rangle_{\textrm{numeric}}$ contains the terms in Eqs. (\ref{eq:vacpolren}),  (\ref{eq:gttren}),  
 (\ref{eq:gphiphiren})  that are in curly brackets, while $\langle \hat{T}^{\mu}{}_{\nu}\rangle_{\textrm{analytic}}$ is comprised of those terms in square brackets. We give explicitly the expression for $\langle \hat{T}^{\mu}{}_{\nu}\rangle_{\textrm{analytic}}$ in Appendix~\ref{app:TAnalytical}. With these definitions, both $	\langle\hat{T}^{\mu}{}_{\nu}\rangle_{\textrm{numeric}}$ and $	\langle\hat{T}^{\mu}{}_{\nu}\rangle_{\textrm{analytic}}$ are independently conserved. In fact  $	\langle\hat{T}^{\mu}{}_{\nu}\rangle_{\textrm{analytic}}$ is equal, up to the addition of a conserved tensor, to the equivalent analytic component obtained by the AHS method \cite{AHS1995}, which in turn reproduces Page's approximation for a massless field. The fact that our direct method reproduces the AHS results, whose derivation depended on the use of the WKB approximation to the radial Green function, is surprising, particularly when contrasted with the results of the extended coordinates approach for a thermal state \cite{taylorbreenottewill:2021}. In that case the equivalent analytic component does not reproduce the AHS equivalent and neither the  analytic nor numerical components are independently conserved, rather only their combination is conserved.

\section{Application to the Reissner-Nordström Space-Time}
\label{sec:ReissnerNordstrom}
In this section, we apply the renormalization scheme presented developed herein to the case of scalar fields in the Boulware state in the Reissner-Nordström black hole space-time:
\begin{align}\label{eq:ReissnerNordström}
	ds^{2}=
	&
	-\frac{\left(r-r_{+}\right)\left(r-r_{-}\right)}{r^{2}}dt^{2}+\frac{r^{2}}{\left(r-r_{+}\right)\left(r-r_{-}\right)}dr^{2}\nonumber\\
	&
	+r^{2}\left(d\theta^{2}+\sin{\theta}^{2}d\phi^{2}\right),
\end{align}
where $r_{\pm}=M\pm\sqrt{M^{2}-Q^{2}}$

We will first verify this new method by reproducing results for the non extremal case, which were obtained via the state subtraction method in \cite{Arrechea:2023}. We then apply the method to the case of an extremal black hole, where we investigate the regularity of the Boulware state on the event horizon for both a massless and massive scalar field. While the regularity of the RSET has previously been examined by Anderson et al. \cite{AndersonEBH}, we are not aware of any previous exact results for a massive field.

Before presenting results for both the non-extremal and extremal cases, we will discuss the numerical implementation of the method for both cases.
\subsection{Numerical Implementation}
\subsubsection{Non-Extremal Modes}
As seen from Eqs.~(\ref{eq:vacpolren}-\ref{eq:gphiphiren}), the mode sums required to compute the RSET can be expressed in terms of derivatives of $\langle\hat{\phi^{2}}\rangle_{\text{ren}}$. In particular, the only mode sum we need to calculate is that for the one-dimensional Green function $g_{\omega l}$~\eqref{eq:Gmodesum}, defined as the normalized product of solutions to the radial equation~\eqref{eq:Euclideanradialeqn}. 

Thus, to obtain the RSET with sufficient accuracy (as, for example, to check its regularity at the horizon) we need to generate highly-precise results for the vacuum polarization and, in turn, for the Euclidean modes. 

For the non-extremal case the computation of the $p_{\omega l}(r)$ modes is simplified by recasting the radial equation, ~\eqref{eq:Euclideanradialeqn}, into a confluent Heun form. This is achieved by expressing the dependent variable $\chi_{\omega l}(r)$ as:
\begin{align*}
    \chi_{\omega l}(r)=e^{-(\omega-\bar{\omega})r}e^{\omega\,r_{*}} Y(r)
\end{align*}
and introducing a new independent variable:
\begin{align*}
    z&=\frac{r_{+}-r}{r_{+}-r_{-}}.
\end{align*}
The radial equation then takes the form of the confluent Heun equation:
\begin{align}
\label{eq:Heun}
	z(z-1) Y''(z)+ (\delta_1 (z-1)+\delta_2 z+z(z-1)\delta_3) Y'(z)\nonumber\\
 	+(q_2 z-q_1)Y(z)=0
\end{align}
where
\begin{align}
	\label{eq:Heunparameters}
	q_1&=l(l+1)+r_{+}^{2}(\omega-\bar{\omega})^{2}-(r_{+}+r_{-})(\omega-\bar{\omega})-2\,r_{+}\bar{\omega}\nonumber\\
	q_2&=(r_{+}^{2}-r_{-}^{2})(\omega-\bar{\omega})^{2}-2(r_{+}-r_{-})\bar{\omega}\nonumber\\
	\delta_1&=1+\frac{2\omega\,r_{+}^{2}}{r_{+}-r_{-}}\nonumber\\
	\delta_2&=1-\frac{2\omega\,r_{-}^{2}}{r_{+}-r_{-}}\nonumber\\
	\delta_3&=-2\bar{\omega}(r_{+}-r_{-})\nonumber\\
	z&=\frac{r_{+}-r}{r_{+}-r_{-}}
\end{align}
and with 
\begin{align}
 \bar{\omega}=\sqrt{\omega^{2}+\mu^{2}}.
\end{align}
If we let $\mathsf{H}(q_1,q_2, \delta_1,\delta_2,\delta_3;z)$ denote the solution to Eq. \eqref{eq:Heun} that is analytic in the vicinity of $z=0$ and normalized to unity there, then we may express the $p_{\omega l}$ solution as:
\begin{align}
	p_{\omega l}(r)&=e^{-(\omega-\bar{\omega})r}e^{\omega\,r_{*}}\mathsf{H}(q_1,q_2,\delta_1,\delta_2,\delta_3;z).
\end{align}
Computing the $q_{\omega l}(r)$ modes is computationally harder. While these modes can still be written in confluent Heun form, the Heun functions with the appropriate boundary conditions for $q_{\omega l}(r)$ are not built into Mathematica or Maple. There are several options one can consider for computing $q_{\omega l}(r)$ but we found it most efficient to simply numerically integrate the radial equation inwards from a large $r$ value using an asymptotic expansion for the initial conditions (see for example \cite{AHS1995} for details of the asymptotic expansion). The initial conditions were optimized so that the asymptotic expansion solved the wave equation to our working precision with the least number of terms in the asymptotic expansion and for the smallest reasonable $r$ value at which this precision could be achieved. Using this approach, the mode solutions and their derivatives that were generated were accurate to at least 30 significant digits. We tested this accuracy by checking the constancy of the Wronskian over the radial grid for the solution pairs $\{p_{\omega l}(r),q_{\omega l}(r)\}$.

\subsubsection{Extremal Modes}
The extremal black hole metric in its Euclideanized form is
\begin{align}
	ds^{2}
	&
	=
	\left(1-\frac{M}{r}\right)^{2}d\tau^{2}+\left(1-\frac{M}{r}\right)^{-2}dr^{2}\nonumber\\
	&
	+r^{2}\left(d\theta^{2}+\sin^{2}\theta d\phi^{2}\right),
\end{align}
for which the mode equation~\eqref{eq:Euclideanradialeqn} amounts to
\begin{align}\label{eq:EuclideanExtremal}
	\Phi''_{\omega l}+
	\frac{2\Phi'_{\omega l}}{r-M}
	-\left[\frac{r^{4}\omega^{2}}{\left(r-M\right)^{4}}+\frac{\mu^{2}r^{2}+l\left(l+1\right)}{\left(r-M\right)^{2}}\right]\Phi_{\omega l}=0.
\end{align}
For convenience, we follow~\cite{Anderson:1990} and adopt the dimensionless coordinate $\eta=r/M-1$, for which the wave equation is
\begin{align}\label{eq:waveeqs}
	\eta^{4}\Phi''_{\omega l}
	&
	+
	2\eta^{3}\Phi'_{\omega l}
	-\left[\omega^{2}M^{2}\left(\eta+1\right)^{4}\right.\nonumber\\
	&
	\left.+l\left(l+1\right)\eta^{2}+\mu^{2}M^{2}\eta^{2}\left(\eta+1\right)^{2}\right]\Phi_{\omega l}=0.
\end{align}
The radial solutions regular at the horizon and at radial infinity obey, respectively, the asymptotic expansions
\begin{equation}
	p_{\omega l}=\eta^{2\omega M}e^{-\omega M/\eta}\sum_{k=0}^{\infty}a_{k}\eta^{k},~\eta\to0
\end{equation}
and 
\begin{equation}
	q_{\omega l}=\eta^{-2\omega M}e^{\omega M/\eta}e^{-\sqrt{\omega^{2}+\mu^{2}}M\eta}\sum_{k=0}^{\infty}b_{k}\eta^{-k-\beta_{0}},~\eta\to\infty.
\end{equation}
Replacing these relations in~\eqref{eq:waveeqs} we find the following recursion relations for the $a_{k},~b_{k}$ coefficients,
\begin{align}
	a_{1}=
	&
	\left[\tilde{\mu}^{2}+l\left(l+1\right)+2\tilde{\omega}\left(\tilde{\omega} -1\right)\right]\frac{a_{0}}{2\tilde{\omega}},\nonumber\\
	a_{2}=
	&
	\left[\tilde{\mu}^{2}+2\tilde{\omega}^2\right]\frac{a_{0}}{2\tilde{\omega} }\nonumber\\
	&
	+\left[\tilde{\mu}^{2}+l\left(l+1\right)+2\tilde{\omega}\left(\tilde{\omega} -3\right)-2\right]\frac{a_{1}}{4\tilde{\omega}},\nonumber\\
	a_{k}
	=
	&
	\left(\tilde{\mu}^{2}+\tilde{\omega}^{2}\right)\frac{a_{k-3}}{2 \tilde{\omega} u}+\left(\tilde{\mu}^{2}+2\tilde{\omega}^{2}\right)\frac{a_{k-2}}{\tilde{\omega} u}\nonumber\\
	&
	+\left[k\left(1-k\right)+
	\tilde{\mu}^{2}+l\left(l+1\right)\right.\nonumber\\
	&
	\left.-4\tilde{\omega} k+2\tilde{\omega} \left(1+\tilde{\omega} \right)\vphantom{M^{2}}\right]\frac{a_{k-1}}{2\tilde{\omega} k},
\end{align}
for the $p_{\omega l}$ mode, where $\tilde{\omega}=\omega M$, $\tilde{\mu}=\mu M$,
with $a_{0}=1$. For the $q_{\omega l}$ mode we have, instead,
\begin{align}
	\beta_{0}
	=
	&
	1-2\tilde{\omega}+\sqrt{\tilde{\omega}^2+\tilde{\mu}^2}+\frac{\tilde{\omega}^2}{\sqrt{\tilde{\omega}^2+\tilde{\mu}^2}},\nonumber\\
	b_{1}
	=
	&
	\left[\tilde{\mu}^2 +l\left(l+1\right)+\beta_{0}\left(1-\beta_{0}-4\tilde{\omega}\right)\right.\nonumber\\
	&
	\left.+2\tilde{\omega} \left(1+\tilde{\omega} -\sqrt{\tilde{\omega}^2+\tilde{\mu}^2}\right)\right]b_{0}\nonumber\\
	&
	\times{\left[2\tilde{\mu}^2-2\beta_{0} \sqrt{\tilde{\omega}^2+\tilde{\mu}^2}+4\tilde{\omega}\left(\tilde{\omega}-\sqrt{\tilde{\omega}^2+\tilde{\mu}^2}\right)\right]^{-1}},\nonumber\\
	b_{k}
	=
	&
	\frac{\left(\beta_{0}+v-2\right)\tilde{\omega}b_{k-2}}{\left(\beta_{0}+k-1\right)\sqrt{\tilde{\omega}^2+\tilde{\mu}^2}-2\tilde{\omega}\left(\tilde{\omega}-\sqrt{\tilde{\omega}^2+\tilde{\mu}^2}\right)-\tilde{\mu}^2}\nonumber\\
	&
	\left[-k^2+k\left(3-2\beta_{0}-4\tilde{\omega}\right)+l\left(l+1\right)+\tilde{\mu}^2-\beta_{0}^2\right.\nonumber\\
	&
	\left.+\beta_{0}\left(3-4\tilde{\omega}\right)+2\tilde{\omega}\left(3+\tilde{\omega}-\sqrt{\tilde{\omega}^2+\tilde{\mu}^2}\right)-2\right]b_{k-1}\nonumber\\
	&
	\times\left[2\left(\beta_{0}+k-1\right)\sqrt{\tilde{\omega}^2+\tilde{\mu}^2}\right.\nonumber\\
	&
	\left.-4\tilde{\omega}\left(\tilde{\omega}-\sqrt{\tilde{\omega}^2+\tilde{\mu}^2}\right)-2\tilde{\mu}^2\right]^{-1},
\end{align}
with $b_{0}=1$.

\subsubsection{Accuracy Issues}
The numerical implementation of the direct method is more challenging than the extended coordinate method developed for thermal states, which uses a discretized frequency spectrum.\\
Most of the issues are related to the sampling of the continuous frequency spectrum. To numerically perform the integral in $\omega$, one needs to interpolate the integrand over a predefined $\omega$-grid. This grid must be fine enough so that the interpolation function works well, ensuring the result is independent of the grid choice. This becomes challenging when the integrand is very large and rapidly decaying. In this context, we found it necessary to push the cutoff $\lambda$ to values no smaller than $1/10$ and to increase the density of the $\omega$-grid immediately after the cutoff. This is done to ensure the interpolation function can properly approximate the rapidly decaying behavior of the cutoff dependent terms in the integrand, which are much larger at smaller $\omega$.\\
Another issue is related to the cancellation of the cutoff-dependent terms. We have shown analytically that the renormalized vacuum polarization does not depend on the choice of the cutoff, as the cutoff-dependent terms ultimately cancel each other. However, achieving this outcome numerically is more challenging and can significantly impact the accuracy of the results. For the case of a thermal state, one can increase the accuracy of the results, by increasing the expansion order of the singular field ($m$ in Eq.~(\ref{eq:hadamard_mode_sum_1})). Moreover, this is extremely efficient numerically as only a very modest number of modes are required. However for the direct approach, as the magnitude of the cutoff-dependent terms increases with increasing order, so does the potential error introduced by the numerical cancellation of these terms. To counteract this, as the expansion order increases, one has to also increase the density in the grid in $\omega$ immediately after the cutoff. In effect if one increases the expansion order, one has to also significantly increase the number of numerical modes required, making the process increasingly inefficient. Hence, there is a balance to be struck between the accuracy of the results and the efficiency of the method. For the results contained in this paper, we have found that a third order expansion ($m=3$) and summing/integrating up to $l=40$/$ \omega=5$ gives sufficiently accurate results, with the conservation equation satisfied to approximately 9-10 decimal places. 

There is an important exception to this choice and that is in the vicinity of the event horizon. All the cutoff-dependent terms vanish on the event horizon and therefore as we approach the horizon the error introduced by these terms also decreases. This is particularly true in the extremal case, where the metric function $f(r)$ has a double root. Therefore, for points close to the horizon (up at a distance of approximately $M/10$) we employ a sixth order expansion to obtain our results and find that the  conservation equation is satisfied to approximately 13-15 decimal places for a massless field and 10-11 decimal places for the massive case.
In addition, particularly near the horizon, the integral is dominated by the low $\omega$ behavior. Hence, it is crucial to select the smallest point in the $\omega$ grid accordingly and extrapolate the integrand to $\omega=0$ to avoid losing important contributions.

Besides the problems related to the continuous $\omega$ spectrum, we have found different representations for $\mathcal{I}_{\omega l}(a,b\,|r)$ and $\mathcal{J}_{\omega l}(a,b\,|r)$. The most amenable to numerical implementation are the ones reported in the derivations above, which involve differences of two hypergeometric functions. However, these representations fail whenever $z=\frac{|\omega|\,r}{\sqrt{f}} \gg 1$, due to catastrophic cancellation. For the region of the domain where \mbox{$z \gg 1$}, the representations in terms of derivatives of modified Bessel functions and generalized incomplete gamma functions are more efficient  and not subject to numerical errors. We include them in appendix \ref{app:other_rep} for convenience.

\subsection{Non-Extremal Results}
\begin{figure}
	\includegraphics[width=0.5\textwidth]{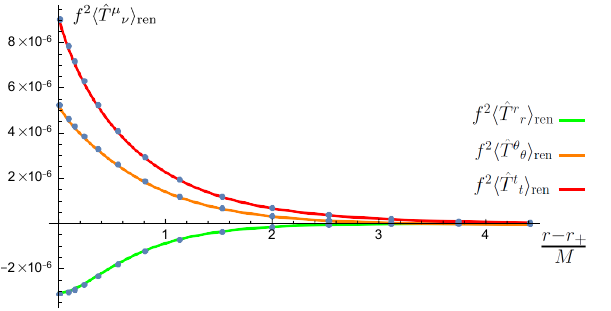}
	\caption{RSET components in the Boulware state for a massless, minimally coupled scalar field calculated via the direct method (line) and the state subtraction method (dots). The background spacetime is the non-extremal Reissner-Nordström spacetime with $Q=2M/10$. }
	\label{fig:compar}
\end{figure}

In Fig.~(\ref{fig:compar}) we compare the results of the direct method outlined in this paper to the corresponding results obtained via a state-subtraction approach in \cite{Arrechea:2023}, for the calculation of the RSET for a massless, minimally coupled scalar field in the Boulware state. The background spacetime on which this calculation is performed is the non-extremal Reissner-Nordström  spacetime with \mbox{$Q=2M/10$}. As is evident from the plot in Fig.~(\ref{fig:compar}), there is good agreement between the two approaches, verifying the new method presented in this paper. However, the state-subtraction method appears to be more accurate, as measured by the conservation equation, in the vicinity of the event horizon. This suggests that for space-times where there exists a Hartle-Hawking state to use as a reference state, taking the state-subtraction approach outlined in \cite{Arrechea:2023} is the appropriate choice. However, for other space-times, such as the extremal Reissner-Nordström spacetime considered in the next section, where we do not have a Hartle-Hawking state to exploit, the direct approach of this paper is required for the RSET calculation for scalar field in the Boulware state.

Returning to the splitting into independently conserved tensors in~\eqref{eq:Tsplit}, we observe that the leading-order divergence of the RSET at the horizon is determined entirely by $\langle\hat{T}^{\mu}{}_{\nu}\rangle_{\textrm{analytic}}\propto f^{-2}$, with $\langle\hat{T}^{\mu}{}_{\nu}\rangle_{\textrm{numeric}}\propto f^{-1}$ amounting to a sub-leading correction. This indicates that $\langle\hat{T}^{\mu}{}_{\nu}\rangle_{\textrm{analytic}}$ indeed captures exactly the singular behaviour of the Boulware state at event horizons, and further justifies its use in backreaction analyses as an analytic approximation to the exact RSET~\cite{Arrechea:2022dvy}. 

\subsection{Extremal Results}
In this section we present the results of the extended coordinates approach to the calculation of the RSET for a scalar field in the Boulware quantum state on an extremal Reissner-Nordström spacetime.

\begin{figure}
	\includegraphics[width=0.5\textwidth]{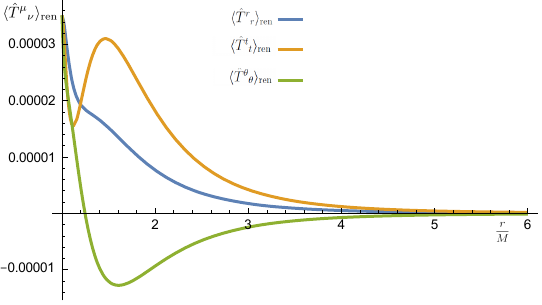}
	\caption{Plot of the components $\langle\hat{T}^{\mu}{}_{\nu}\rangle_{\textrm{ren}}$ for a conformally coupled massless field in the Boulware state, on an extremal Reissner-Nordström spacetime.}
	\label{fig:extremalm0full}
\end{figure}

\begin{figure}
	\includegraphics[width=0.5\textwidth]{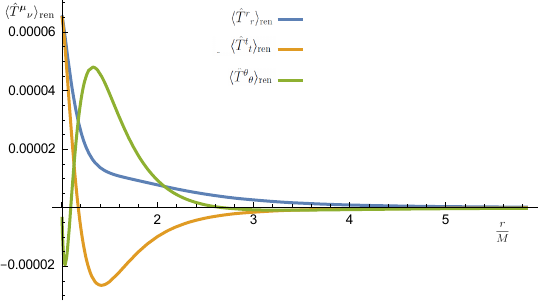}
	\caption{Plot of the components $\langle\hat{T}^{\mu}{}_{\nu}\rangle_{\textrm{ren}}$  for a minimally coupled field with mass $\mu M=1/10$ in the Boulware state, on an extremal Reissner-Nordström spacetime.}
	\label{fig:extremalm1o10full}
\end{figure}

In Figs.~(\ref{fig:extremalm0full}) and (\ref{fig:extremalm1o10full}) we present the components of the RSET in the exterior region of the black hole for a massless and massive field  ($\mu M=1/10$) respectively. For the massless case, on the horizon $\langle\hat{T}^{\mu}{}_{\nu}\rangle_{\textrm{numeric}}$ vanishes and the RSET is given precisely by $\langle\hat{T}^{\mu}{}_{\nu}\rangle_{\textrm{analytic}}$ with 
\begin{equation}
	\langle\hat{T}^{\mu}{}_{\nu}\rangle|_{r=M}=\langle\hat{T}^{\mu}{}_{\nu}\rangle_{\textrm{analytic}}|_{r=M}=\frac{1}{2880 \pi^2 M^4} \delta^{\mu}{}_{\nu},\label{Eq:RSETExtrHor}
\end{equation}
which is equal to the corresponding RSET in the Bertotti-Robinson spacetime \cite{AndersonEBH}. The fact that  $\langle\hat{T}^{\mu}{}_{\nu}\rangle_{\textrm{analytic}}$ is exact on the horizon, suggests that it might serve as a good approximation for the full RSET, at least in the massless case. However this is not the case, with $\langle\hat{T}^{\mu}{}_{\nu}\rangle_{\textrm{analytic}}$ failing to capture the main features of the full results, except in the immediate vicinity of the horizon. Finally, as $r$ increases we see that all RSET components tend to 0, as would be expected for a zero-temperature state.

\begin{figure}
	\includegraphics[width=0.5\textwidth]{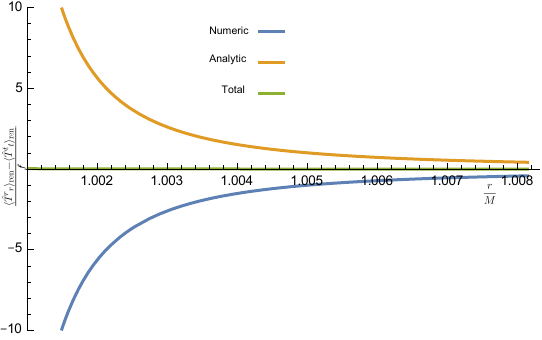}
	\caption{Plot of the cancellation of the divergence in the numerical and analytical contributions to the quantity $\langle\hat{\mathcal{E}}\rangle$ for a minimally coupled field with mass $\mu=M/10$, leaving a total result that appears to be finite on the event horizon of the extremal Reissner-Nordström  spacetime}
	\label{fig:extremalm1o10}
\end{figure}

For the massive case, the situation is quite different. Here both $\langle\hat{T}^{\mu}{}_{\nu}\rangle_{\textrm{analytic}}$ and $\langle\hat{T}^{\mu}{}_{\nu}\rangle_{\textrm{numeric}}$ diverge logarithmically on the horizon, however these divergences cancel to render each of the components of $\langle\hat{T}^{\mu}{}_{\nu}\rangle$ finite there. Specifically, the analytic terms become, in the near-horizon limit,
\begin{align}
\langle\hat{T}^{\mu}{}_{\nu}\rangle_{\textrm{analytic}}|_{r=M}=
&
\frac{\mu^{2}}{8\pi^2}\left[\frac{\left(\xi-\frac{1}{6}\right)}{M^{2}}\text{diag}\left(1,1,-1,-1\right)\right.\nonumber\\
&
\qquad\left.+\frac{\mu^{2}}{4}\delta^{\mu}{}_{\nu}\right]
\log{\left(\frac{r}{M}-1\right)}.
\end{align}
In contrast to the massless case, only the $\langle\hat{T}^{r}{}_{r}\rangle$ and $\langle\hat{T}^{t}{}_{t}\rangle$ components agree on the horizon with $\langle\hat{T}^{\theta}{}_{\theta}\rangle$ taking on a distinct value. For massive fields, the $r \to \infty$ behaviour of the RSET components depends on the renormalisation lengthscale $\ell$. In fact, from the numerical results we find that the RSET approaches the following value:
 \begin{equation}
		\langle\hat{T}^{\mu}{}_{\nu}\rangle_{\infty}=\frac{\delta^{\mu}{}_{\nu}\mu^{4}}{128\pi^{2}}\left[3+4\log\left(\frac{2e^{-\gamma_{\subE}}}{\mu\, \ell} \right)\right].
	\end{equation}
   Therefore, there is a natural choice of the lengthscale for massive fields for which the RSET tends to 0 as $r \to \infty$. To choice is given by:
\begin{align}\label{Eq:Lengthscale}
    \ell=\frac{2}{\mu} \mbox{exp}\{3/4-\gamma_{\subE}\},
\end{align}
and is the value chosen for the results presented in Figs.~(\ref{fig:extremalm1o10full}) and (\ref{fig:extremalm1o10}). For the massless case, there is no such natural choice, so we simply set $\ell=M$.

\begin{figure}
	\includegraphics[width=0.5\textwidth]{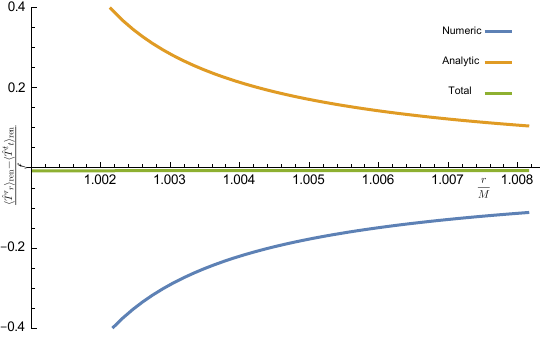}
	\caption{Plot of the cancellation of the divergence in the numerical and analytical contributions to the quantity $\langle\hat{\mathcal{E}}\rangle$ for a minimally coupled massless field, leaving a total result that appears to be finite on the event horizon of the extremal Reissner-Nordström  spacetime }
	\label{fig:extremalm0}
\end{figure}

In Figs.~(\ref{fig:extremalm0}) and ~(\ref{fig:extremalm1o10}) we focus on the near horizon region and investigate the regularity of the Boulware state in an extremal Reissner-Nordström spacetime for a massless and massive field ($\mu M=1/10$) respectively. The quantity  $\langle\hat{T}^{\mu}{}_{\nu}\rangle_{\textrm{analytic}}$ is not regular on the event horizon resulting in an infinite energy density there as perceived by a freely falling observer:
\begin{align}\label{Eq:EDFreeFall}
	\langle\hat{\mathcal{E}}\rangle_{\textrm{analytic}}
 &
 =
\frac{\langle\hat{T}^{r}_{~r}\rangle_{\textrm{analytic}}-\langle\hat{T}^{t}_{~t}\rangle_{\textrm{analytic}}}{f}\nonumber\\
    &
    =-\frac{\mu^{2}\left[3M^{2}\mu^{2}+12\left(\xi-\frac{1}{6}\right)\right]}{96\pi^{2}M^{2}}\left(\frac{r}{M}-1\right)^{-2}\nonumber\\
    &
    \quad+\frac{15M^{4}\mu^{4}-2}{240\pi^{2}M^{4}}\left(\frac{r}{M}-1\right)^{-1}\nonumber\\
    &
    \quad -\frac{1}{60\pi^{2}M^{4}}\log\left(\frac{r}{M}-1\right)+\mathcal{O}{\left(\frac{r}{M}-1\right)}^{0}.
\end{align}
This quantity possesses a $(r-M)^{-1}$ divergence for a massless field and a $(r-M)^{-2}$ for a massive field. 
In Figs.~(\ref{fig:extremalm0}) and ~(\ref{fig:extremalm1o10}), we show that this divergence appears to be precisely canceled by a corresponding divergence from  $\langle\hat{T}^{\mu}{}_{\nu}\rangle_{\textrm{numeric}}$, leaving a total  	$\langle\hat{\mathcal{E}}\rangle$ that is finite on the event horizon. Of course, to prove this definitively would require a closed form expression for the quantity $\langle\hat{\mathcal{E}}\rangle$ on the horizon (or equivalently a near horizon expression for $\langle\hat{T}^{\mu}_{~\nu}\rangle$ to $\mathcal{O}(r-M)^2$). However, we feel the numerical results presented here provide compelling evidence for the regularity of the Boulware state. 

We have therefore used the novel method presented in the paper to confirm the results of \cite{AndersonEBH} for a massless field and provide, to the best of the authors' knowledge, the first exact results for the RSET of a massive scalar field in the extremal Reissner-Nordström  spacetime.

\section{Backreaction in extremal black holes}
\label{sec:backreaction}
Computing the RSET in the background of an extremal black hole naturally begs the question of whether extremal horizons are stable under quantum corrections. For the static, spherically symmetric extremal black holes under consideration, despite the field being in a static zero-temperature vacuum state and thus not inducing any evaporation process at lowest order in the semiclassical approximation, the quantum fields generate a RSET which is (generically) non-vanishing at the horizon and will backreact on the background metric. 

We start by considering the semiclassical Einstein equations,
\begin{equation}
   \label{eq:semiclassicaleqs} G^{-1}\mathsf{G}_{\mu\nu}+\Lambda\,\mathsf{g}_{\mu\nu}+\alpha_{1}\,\mathsf{H}^{(1)}_{\mu\nu}+\alpha_{2}\,\mathsf{H}^{(2)}_{\mu\nu}=8\pi\left(\mathsf{T}_{\mu\nu}^{\textrm{(cl)}}+\epsilon\,\langle \hat{\mathsf{T}}_{\mu\nu}\rangle\right)
\end{equation}
where  $\mathsf{T}_{\mu\nu}^{\textrm{(cl)}}$ is the classical stress-energy tensor, $\langle \hat{\mathsf{T}}_{\mu\nu}\rangle$ are the contributions to the stress-energy due to the quantum fields, and $\mathsf{H}^{(1)}_{\mu\nu}$, $\mathsf{H}^{(2)}_{\mu\nu}$ are the geometrical tensors defined by
\begin{align}
    \label{eq:Htensors}
    \mathsf{H}^{(1)}_{\mu\nu}&=-2\Box \mathsf{R}_{\mu\nu}+\tfrac{2}{3}\nabla_{\mu}\nabla_{\nu}\mathsf{R}+\tfrac{1}{3}\mathsf{g}_{\mu\nu}\Box \mathsf{R}-\tfrac{1}{3}\mathsf{g}_{\mu\nu}\mathsf{R}^{2}\nonumber\\
	&+\tfrac{4}{3}\mathsf{R}\,\mathsf{R}_{\mu\nu}+(\mathsf{R}_{\lambda\rho}\mathsf{R}^{\lambda\rho})\mathsf{g}_{\mu\nu}-4 \mathsf{R}_{\mu\lambda\nu\rho}\mathsf{R}^{\lambda\rho},\nonumber\\
    \mathsf{H}_{\mu\nu}^{(2)}
	&=2\nabla_{\mu}\nabla_{\nu}\mathsf{R}-2\mathsf{g}_{\mu\nu}\Box \mathsf{R}+\tfrac{1}{2}\mathsf{R}^{2}\mathsf{g}_{\mu\nu}-2 \mathsf{R}\,\mathsf{R}_{\mu\nu}.
\end{align}
The tensors in these equations depend on an unknown metric $\mathsf{g}_{\mu\nu}$. Note that we have reinstated Newton's constant $G$ and expressed it on the left-hand side of the equations, hence $\Lambda$ here has the interpretation of the cosmological constant divided by Newton's constant. The $\epsilon$ on the right-hand side is to keep track of quantum corrections and so $\mathcal{O}(\epsilon)\sim\mathcal{O}(\hbar)$. As already mentioned, the renormalization of $\langle \hat{\mathsf{T}}_{\mu\nu}\rangle$ on the right-hand side is ambiguous with a freedom to add terms proportional to the geometrical tensors on the left-hand side of (\ref{eq:semiclassicaleqs}) above. Hence, different renormalizations of the quantum stress-energy tensor are degenerate with different renormalizations of the constants $\alpha_{1}$, $\alpha_{2}$, $\Lambda$ and  $G^{-1}$~\cite{Sanders:2020osl}.

Now, we wish to solve the semiclassical equations perturbatively. We first assume the RSET ambiguity is fixed so that all the freedom is contained in the constants on the left-hand side. In particular, we take the arbitrary renormalization lengthscale $\ell$ as in Eq.~(\ref{Eq:Lengthscale}). Then we  expand the metric as $\mathsf{g}_{\mu\nu}=g_{\mu\nu}+\epsilon\,h_{\mu\nu}$. To keep track of the corrections to the constants coming from the RSET, we write the constants as $\alpha_{1}+\epsilon\delta\alpha_{1}$ etc., where now $\alpha_{1}$ is considered to be the background value. Since we wish to expand perturbatively  about the extremal Reissner-Nordstr{\"o}m solution, we assume $\Lambda$, $\alpha_{1}$ and $\alpha_{2}$ vanish and hence the zeroth order equation is $G^{-1}G_{\mu\nu}=8\pi\, T_{\mu\nu}^{(\textrm{cl})}$, where we reiterate that both sides of this equation are evaluated on the background spacetime. Taking the classical stress-energy tensor to be that derived from the Faraday tensor and assuming spherical symmetry yields the Reissner-Nordstr{\"o}m solution at the zeroth order. Then at first order in $\epsilon$, we have
\begin{align}
   \label{eq:perturbationeqn}
   \delta G_{\mu\nu}+\delta G^{-1}\,G_{\mu\nu}+\delta\Lambda\,g_{\mu\nu} +\delta\alpha_{1}\,H^{(1)}_{\mu\nu}=8\pi\,\langle \hat{T}_{\mu\nu}\rangle_{\textrm{ren}}, 
\end{align}
where the $H^{(2)}_{\mu\nu}$ tensor defined by (\ref{eq:Htensors}) vanishes on the background spacetime. The perturbation of the Einstein tensor is given explicitly by
\begin{align}
  \label{eq:perturbationdef}
  -2\,\delta G_{\mu\nu}=\Box \bar{h}_{\mu\nu}+g_{\mu\nu}\nabla^{\lambda}\nabla^{\rho}\bar{h}_{\lambda\rho}-2\nabla^{\lambda}\nabla_{(\mu}\bar{h}_{\nu)\lambda}\nonumber\\
	-g_{\mu\nu}R^{\lambda\rho}\bar{h}_{\lambda\rho}+R\,\bar{h}_{\mu\nu}
\end{align}
and where $\bar{h}_{\mu\nu}=h_{\mu\nu}-\tfrac{1}{2}g_{\mu\nu}g^{\lambda\rho}h_{\lambda\rho}$ is the trace-reversed metric perturbation. The derivatives in (\ref{eq:perturbationdef}) are with respect to the background metric $g_{\mu\nu}$. The solution space to (\ref{eq:perturbationeqn}) is very large so we focus on the simplest case for which the constants $\delta\Lambda$, $\delta \alpha_{1}$ and $\delta G^{-1}$ either vanish or are higher order in the perturbative expansion so that they do not contribute in (\ref{eq:perturbationeqn}). We can weaken the assumption that $\delta G^{-1}$ vanishes and instead consider that this term can be absorbed into a redefinition of the charge of the background solution, a fact which can be easily seen from the zeroth order equation.

Since we are considering a real quantum scalar field, the electromagnetic stress-energy tensor is not affected by semiclassical contributions~\cite{AndersonEBH}. Following~\cite{York1984},  we assume the full metric coefficients have the form
\begin{align}
     \mathsf{g}_{tt}=-e^{2\psi(r)}\left(1-2m(r)/r\right),\quad \mathsf{g}_{rr}=\left(1-2m(r)/r\right)^{-1},
\end{align}
where $m(r)$ denotes the Misner-Sharp mass, and then we expand these metric functions as
\begin{align}
    \psi(r)\simeq\log{\left[1+\epsilon \mathcal{R}(r)\right]},\quad m(r)\simeq M\left(1-\frac{M}{2r}\right)\left[1+\epsilon \mathcal{M}(r)\right],
\end{align}
with $\epsilon=\hbar/M^2$. To linear order in $\epsilon$, we write the semiclassical Einstein equations in the form
\begin{align}
    \frac{2\left(z-1\right)^2\mathcal{R}'(z)}{M^2 z^3}=
    &
    ~8\pi M^2 \left(\langle\hat{T}^{r}_{~r}\rangle-\langle\hat{T}^{t}_{~t}\rangle\right),\nonumber\\
    -\frac{\mathcal{M}(z)}{M^2 z^4}-\frac{\left(2z-1\right)\mathcal{M}'(z)}{M^2 z^3}=
    &
    ~8\pi M^2 \langle\hat{T}^{t}_{~t}\rangle,\label{Eq:Perturbed}
\end{align}
where we have used the dimensionless coordinate \mbox{$z=r/M$}, and we have dropped the subscript ``ren'' on the RSET components for typographical convenience. The first equation is clearly the difference between the radial and temporal equations, and the angular equations are such that the Bianchi identities are satisfied. Once the RSET is specified (here we replace it by the exact, numerical values for massless and massive fields shown in the previous section), we can integrate Eqs.~\eqref{Eq:Perturbed} for $\mathcal{M}$ and $\mathcal{R}$. The full solutions for the perturbations are numeric and we can only calculate them within the region for which we have numerical RSET values at our disposal. 

For massless fields in the near-horizon limit $z\to z_{\text{e}}$, we can make use of the fact that the energy density and pressures of the exact RSET are given, to leading order in $z-z_{\text{e}}$, by~\eqref{Eq:RSETExtrHor}
to obtain the following analytic solutions for the perturbations
\begin{equation}
    \mathcal{M}=-\frac{C_{0}}{1080\pi}+\mathcal{O}\left(z-z_{\text{e}}\right),\quad \mathcal{R}=\frac{K_{0}}{1080\pi}+\mathcal{O}\left(z-z_{\text{e}}\right),
\end{equation}
where $\left\{C_{0},K_{0}\right\}$ are arbitrary integration constants. For massive fields, however, we cannot follow a similar argument since we ignore the exact values the RSET must reach at the extremal horizon, and extrapolating our numerical results introduces spurious errors. However, backreaction analyses have been done in the limit of large field mass~\cite{Matyjasek:2001yf}.
Written in ingoing Eddington-Finkelstein coordinates, the metric components become
\begin{align}
    g_{vv}=
    &
    g_{tt}=-\frac{\epsilon C_{0}}{1080\pi}+\mathcal{O}\left(z-z_{\text{e}}\right),\nonumber\\
    g_{vr}=
    &
    e^{\psi}=1+\frac{\epsilon K_{0}}{1080\pi}+\mathcal{O}\left(z-z_{\text{e}}\right).
\end{align}
In light of these expressions, we observe that the metric receives a correction so that $g_{tt}$ is no longer vanishing at $z=z_{\text{e}}$. For negative $\epsilon C_{0}$ we have $g_{tt}>0$, hence the surface $z=z_{\text{e}}$ appears to lie inside a trapped region. In this case perturbative corrections have shifted the position of the horizon from its classical value $z_{\text{e}}=1$ to a new position $z_{\rm H}>z_{\text{e}}$ which must be found numerically. The surface gravity associated to this quantum-corrected horizon is equal to
\begin{equation}
    \kappa_{\rm H}=\frac{e^{\psi}}{2M}\frac{\partial}{\partial z}\left(1-\frac{2Mm}{z}\right)+\frac{e^{\psi}}{M}\left(1-\frac{2Mm}{z}\right)\frac{\partial\psi}{ \partial z}\bigg|_{z_{\rm H}}.
\end{equation}

Extremal horizons are characterized by a vanishing surface gravity, i.e., a quadratic root in $g_{vv}$. In the Reissner-Nordström metric, this comes from the fact that $Q=M$. The RSET makes the Misner-Sharp mass acquire a radial dependence, but does not affect the charge, allowing for the possibility of breaking the delicate balance of scales required by extremal horizons. In practice what we observe is that the backreacted spacetime, locally around $z=z_{\text{e}}$, becomes like the sub-extremal or super-extremal Reissner-Nordström spacetimes depending on which sign we choose for the mass perturbations. Since we do not have exact results for $z<z_{\text{e}}$, we cannot obtain the quantum-corrected metric in the region $z\ll z_{\rm H}$.

\begin{figure}
    \centering\includegraphics[width=\linewidth]{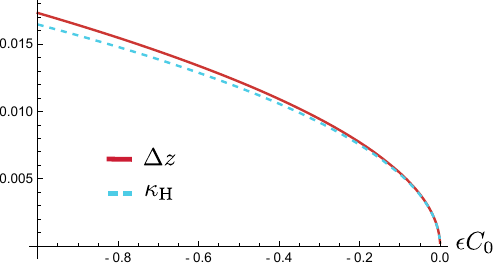}
    \caption{Position of the perturbed horizon $\Delta z=z_{\rm H}-z_{e}$ and its surface gravity $\kappa_{\rm H}$ for a semiclassically corrected extremal black hole. Corrections from minimally coupled massive and massless fields shift the Misner-Sharp mass by a constant value, which either breaks the extremal horizon into a pair of inner and outer horizons (if \mbox{$C_{0}<0$}), or eliminates any horizon entirely (if \mbox{$C_{0}>0$}).
    Our results do not change qualitatively as long as $\epsilon K_{0}\ll 1$, so we set $K_{0}=0$ here.}
    \label{Fig:Perturb}
\end{figure}

Together with shifting the position of the horizon outwards, taking  \mbox{$\epsilon C_{0}<0$} makes it acquire a positive surface gravity. The relative position and surface gravity of the quantum-corrected horizon are shown in Fig.~\ref{Fig:Perturb}. For massless fields, we see this non-extremal horizon smoothly tends to a horizon of the extremal kind in the $\epsilon C_{0}\to0$ limit. In the massive case, since we lack RSET values arbitrarily close to $z_{\text{e}}$, we have not been able to generate reliable results in the $\epsilon C_{0}\to0$ limit. Nonetheless, for large and negative $\epsilon C_{0}$ values, the curves for the massless and massive cases are indistinguishable. 

We can integrate Eqs.~\eqref{Eq:Perturbed} away from the black hole to obtain the behaviour of pertubations at large $z$. The $\mathcal{R}$ function can be absorbed in a redefinition of the $t$ coordinate, while the $\mathcal{M}$ function shifts the  Arnowitt–Deser–Misner (ADM) mass by a constant factor. Under the choice of renormalized constants \mbox{$\delta\Lambda=\delta G^{-1}=0$}, the RSET decays at large $z$ as \mbox{$\langle \hat{T}^{\mu}_{~\nu}\rangle\propto \mathcal{O}\left({z}^{-4}\right)$} for massive fields and as \mbox{$\langle \hat{T}^{\mu}_{~\nu}\rangle\propto \mathcal{O}\left({z}^{-6}\right)$} for massless fields. In both cases, it can be shown from~\eqref{Eq:Perturbed} that, for $z\gg z_{e}$, \mbox{$\mathcal{M}=C_{\infty}+\mathcal{O}\left({z}^{-1}\right)$}. 

\begin{figure}
    \centering
    \includegraphics[width= \linewidth]{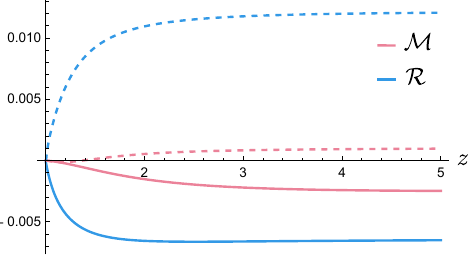}
    \caption{Solutions for the perturbations $\mathcal{R}(r)$ of the redshift function  (in blue) $\mathcal{M}(r)$  and the Misner-Sharp mass (in pink) in terms of the radial coordinate $z$. The continuous lines correspond to the $\mu M =0$ case and the dashed lines to the $\mu M=1/10$ case. Taking $C_{0}=K_{0}=0$, we have integrated  
    within the domain of $z$ values in which we have reliable numerical results for the RSET.
    }
    \label{Fig:PerturbSols}
\end{figure}

Figure~\ref{Fig:PerturbSols} shows the $\mathcal{R}$ and $\mathcal{M}$ functions for the minimally coupled massless field (continuous lines) and the minimally coupled $\mu M=1/10$ field. We have integrated the equations in the radial domain for which we have reliable results for the RSET. These are $z\in[M, 5.75M]$ and $z\in[1.005M,5.75M]$ in the massless and massive cases, respectively. For the massless case, preserving the extremal horizon $(C_{0}=0)$ translates into a negative correction to the ADM mass of $C_{\infty}\approx-2.5\times 10^{-3}$, so that $M_{\text{ADM}}=M+C_{\infty}<Q$. In this case, in order to preserve an extremal horizon, we would need to start with a sub-extremal black hole and perturb it semiclassically.
For the massive case, integrating outwards with $C_{0}=0$ we obtain $C_{\infty}\approx9.7\times10^{-4}$, so that $M_{\text{ADM}}>Q$. In this case extremality would be preserved if we started with a super-extremal black hole and corrected it semiclassically. This discrepancy in $\mathcal{M}$ at large distances can be traced back to the differing signs of the $\langle\hat{T}^{t}_{~t}\rangle$ component in each case (see Figs.~\ref{fig:extremalm0full} and~\ref{fig:extremalm1o10full}). For massless fields, the $\xi$-dependent contributions to the RSET vanish at the horizon and as $r\to\infty$. Regions in which $\langle\hat{T}^{t}_{~t}\rangle$ changes sign only appear for $\xi\gtrsim 1/4$, hence these results are expected to hold qualitatively for a broad range of curvature couplings.

So far we have considered $\alpha_{1}$=0. For non-vanishing $\alpha_{1}$ in Eq.~\eqref{eq:perturbationeqn}, we observe the $H^{(1)}_{\mu\nu}$ tensor is sub-leading with respect to the RSET as the event horizon of the classical solution is approached, hence it does not modify the position of the perturbed horizon nor its surface gravity. For large $z$ this tensor decays $\propto z^{-6}$, hence the perturbations 
$\mathcal{M}$ and $\mathcal{R}$ still approach constant values at infinity. However, the contribution from  $H^{(1)}_{\mu\nu}$ for intermediate $z$ values can change the sign of the perturbations $\mathcal{M}$ and $\mathcal{R}$ at infinity. For example, a sufficiently large and negative $\alpha_{1}$ leads to a positive $C_{\infty}$ in the massless minimally coupled case.

The loss of extremality is expected to occur
due to a breakdown of the delicate balance between charge and mass contributions that extremal horizons require.
Since we are only incorporating the vacuum polarization effects from an uncharged massless scalar field, this balance is broken at first order in perturbation theory. However, it is still plausible that the vacuum polarization effects from the electromagnetic field only displace the position of the horizon without changing its extremal nature.

On the one hand, for negative $\epsilon C_{0}$, it is clear (since the corrections are perturbative) that we are observing a splitting of the extremal horizon into a pair of outer and inner horizons. 
By solving Eq.~\eqref{eq:semiclassicaleqs} perturbatively to higher order in $\hbar$, we would find: i) that the vacuum state in which this RSET is evaluated is no longer a zero-temperature state, and ii) that this state is singular at the newly formed inner horizon~\cite{Zilbermanetal2022}.  
On the other hand, for positive $\epsilon C_{0}$ the function $g_{vv}$ has zero roots, hence the extremal horizon disappears entirely and the perturbed metric describes a horizonless (and singular) object, whose associated vacuum state has zero temperature. Finally, between both regimes there is a separatrix solution with $C_{0}=0$ that preserves both the position and the surface gravity of the extremal horizon. This solution could potentially exist for all orders in $\hbar$ only if we fine-tune the perturbations so that their contribution to the Misner-Sharp mass vanishes at every order in $\epsilon$. Of course, the treatment adopted in this section only deals with static solutions from the start, hence we remain agnostic in regard to the stability of extremal horizons under time-dependent perturbations.

\section{Discussion and Conclusions}
\label{sec:conclusions}
In this paper, we have provided a generalisation of the extended coordinate prescription to directly compute the
renormalized stress-energy tensor (RSET) of scalar fields in the Boulware vacuum. While the extended coordinates approach was first developed for the calculation of  expectation values for a scalar field in the Hartle-Hawking state, it can be used to obtain results for other quantum states using a state subtraction approach \cite{Arrechea:2023}. However, for spacetimes where one does not have a Hartle-Hawking state to leverage as a reference state, such as an extremal black hole or the spacetime of a star then a direct approach to compute the RSET is required.

In the extended coordinate approach we take an expansion of the
Hadamard parametrix in extended coordinates, which encodes the short-distance
behaviour of the two-point function, and we rewrite the
terms in this expansion as mode-sums of the same form as those in the unrenormalized Green function, which may then be renormalized in a mode-by-mode manner. The main technical challenge in applying this approach to a scalar field at zero temperature arises from having a continuous frequency spectrum and hence the mode sums involve integrals over $\omega$ (as opposed to the Hartle-Hawking case where we sums over discrete values of $\omega$). As result, when we express terms in the Hadamard
parametrix as mode-sums, we find that the integrands
are not ordinary functions of $\omega$ and we must treat these mode-sums rigorously as generalized functions. When we do this, we see that an infra-red cutoff naturally appears in the integrals over $\omega$ in such a way so that the final results are independent of this cut-off. 

Having developed this approach, we first applied it to the non-extremal Reissner-Nordström spactime. In this case we reproduced the results obtained via the state subtraction approach in \cite{Arrechea:2023}, thereby verifying the new direct approach. We then applied the direct approach the the calculation of the RSET in an extremal Reissner-Nordström spacetime, for both a massless and massive scalar field. The massless case had previously been considered in \cite{AndersonEBH}, however, to the best of the authors' knowledge, the results in this paper are first results obtained for a massive field.  

Once we have calculated the RSET components we used these results to investigate the regularity of the Boulware state by calculating the energy density on the horizon as perceived by a freely falling observer. Here we find strong numerical evidence that the Boulware state is regular for both the massless and massive case. However, we stress that these results are indicative only and to say anything conclusive requires exact results on the horizon. We hope to report on this in the near future. 

Despite lacking a definite proof of the RSET regularity at extremal horizons, we can nonetheless assume that it is indeed regular to give a hint on the backreaction effects it would entail. By solving the static semiclassical equations perturbatively in $\hbar$~\cite{York1984} we have seen that, unless perturbations are fine-tuned to every order in the expansion, extremal horizons are unstable. If these perturbations contribute positively to the Misner-Sharp mass, the extremal horizon splits into a pair of outer and inner horizons, while in the opposite scenario the horizon disappears completely, leaving no marginally trapped surfaces in the spacetime. We also analyzed how extremality at the horizon results in a change in the ratio between the ADM mass and the charge, which depends on the overall sign of the semiclassical energy density.

The generalization of the extended coordinate method to the Boulware state opens the window towards obtaining the RSET of scalar fields in previously unexplored situations, particularly, in stellar spacetimes. 
Recent works that considered perfect fluid stars show indications that the RSET becomes a dominant contribution in stellar interiors as these stars approach their maximum compactness limits~\cite{Hiscock1988,Reyes2023,Arrecheaetal2023}. Since previous works were based on different RSET approximations, a calculation of the exact RSET will allow to address their validity. These explorations have great implications for the theoretical plausibility of ultracompact stars sourced by semiclassical effects.

\acknowledgments{The authors thank the Institute for Fundamental Physics of the Universe for hosting their visit under the Team Research program \textit{Dark Compact Objects and The Quantum Vacuum}.
JA acknowledges funding from the Italian Ministry of Education and Scientific Research (MIUR) under the grant
PRIN MIUR 2017-MB8AEZ. \\
LP acknowledges funding from Taighde Éireann - Research Ireland under Grant number GOIPG/2024/4003.\\
The authors would like to thank Marc Casals for helpful conversations.}
\begin{widetext}
\section*{Appendix}
\appendix
\section{Cut-Off Independence}
\label{app:cutoff}
There is a subtle cancellation happening between the terms dependent on $\lambda$ in $\mathcal{J}_{\omega l}(a,b\,|r)$ and the finite terms in \eqref{eq:hadamard_mode_sum_2} which also depend on $\lambda$. We can make this cancellation explicit by analyzing the terms in the renormalized vacuum polarization that contain the cutoff, which we refer to as $\mathsf{w}^{(\lambda)}$:
\begin{align}
\label{eq:phi_sq_cutoff}
    \mathsf{w}^{(\lambda)}&=\frac{\mathcal{T}_{00}^{(\textrm{l})}(r)}{8\pi^{2}}\Bigg[\int_{-\infty}^{\infty}d\omega\,\theta\left(\omega^{2}-\lambda^{2}\right)|\omega|^{-1}
    +\log(\lambda^2)\Bigg]\nonumber\\
    &+\frac{\sqrt{\pi}}{8\pi^{2}}\sum_{a=1}^{m-1}\sum_{b=0}^{a}\mathcal{T}_{ab}^{(\textrm{l})}(r)\frac{f^{a-b}2^{2a}}{\Gamma(-a+\tfrac{1}{2})}\,\Bigg[a!\int_{-\infty}^{\infty}d\omega\,\theta\left(\omega^{2}-\lambda^{2}\right)|\omega|^{-2a-1}
    -\frac{(a-1)!}{\lambda^{2a}}\Bigg].
\end{align}
We have separated the $\mathcal{T}_{00}^{(l)}(r)$ and $a>0$ terms in $\mathcal{J}_{\omega l}(a,b\,|r)$ and used the identity \eqref{eq:legendreidentity} to truncate the $l$ sum as well as introduced $k=0$, only condition for these terms being nonzero at coincidence.
We can show that the renormalized vacuum polarization is independent on the choice of the cutoff by verifying that
\begin{align}
    \frac{\partial}{\partial\lambda}\left(\mathsf{w}^{(\lambda)}\right)
    =\frac{\mathcal{T}_{00}^{(\textrm{l})}(r)}{8\pi^{2}}\left(-\frac{2}{\lambda}+\frac{2}{\lambda}\right)
    +\frac{\sqrt{\pi}}{8\pi^2}\sum_{a=1}^{m-1}\sum_{b=0}^{a}\mathcal{T}_{ab}^{(\textrm{l})}(r)\frac{f^{a-b}2^{2a}}{\Gamma(-a+\tfrac{1}{2})}\left(-2 \,a!\lambda^{-2a-1}+2 \,a!\lambda^{-2a-1}\right)=0,
\end{align}
where we made use of
\begin{equation}
    \frac{\partial}{\partial\lambda}\left(\theta\left(\omega^{2}-\lambda^{2}\right)\right)=-\left(\delta\big(\omega+\lambda\right)+\delta\left(\omega-\lambda\right)\big).
\end{equation}
Similarly, it can be demonstrated that the RSET is also completely independent of the choice of the cutoff.

\section{Other representations for direct and tail regularization parameters}
\label{app:other_rep}
In this appendix, we report different but equivalent representations for the functions $\mathcal{I}_{\omega l}(a,b\,|r)$ and $\mathcal{J}_{\omega l}(a,b\,|r)$. Above, we have presented the derivation for these expressions in terms of hypergeometric functions. The alternative representations below involve the modified Bessel functions $I_{\nu}(z)$ and $K_{\nu}(z)$ and the generalized incomplete gamma function $\Gamma(a,z_0,z_1)=\int_{z_0}^{z_1}t^{a-1}e^{-t}dt$.
\begin{align}
    \mathcal{I}_{\omega l}(a,b\,|r)&=
    \frac{r^{2 b+1}f^{-b-1/2}}{\left(2^b b!\right) \left(2 r^2\right)^{b+1}} \sum _{p=0}^b \frac{(-1)^a 2^p (p)_{2 b-2 p}}{\Gamma (b-p+1)}\frac{\partial ^{2 a+p}}{\partial |\omega| ^{2 a+p}}\left(|\omega| ^p I_{l+\frac{1}{2}}\left(\frac{|\omega|  r}{\sqrt{f}}\right) K_{l+\frac{1}{2}}\left(\frac{|\omega|  r}{\sqrt{f}}\right)\right)\nonumber\\
    &=\frac12\frac{(-1)^a}{2^{2b} b!  (f)^{b+1/2}r} \sum _{p=0}^b \frac{ 2^p (p)_{2 b-2 p}}{\Gamma (b-p+1)}\sum_{k=0}^{2a+p}\binom{2a+p}{k}\frac{p!\,|\omega|^{-2a+k}}{(-2a+k)!}\sum_{d=0}^{k}\binom{k}{d}\left(\frac{r}{\sqrt{f}}\right)^k\frac{(-1)^d}{2^k}\nonumber\\
    &\quad\times\sum_{c=0}^{2k-2d}\binom{k-d}{c}I_{l+1/2-k+d+2c}\left(\frac{|\omega|  r}{\sqrt{f}}\right)\sum_{n=0}^{2d}\binom{d}{n}K_{l+1/2-d+2n}\left(\frac{|\omega|  r}{\sqrt{f}}\right)\nonumber\\
    &=\lim_{x\to-1}\frac{2^l (-1)^{a+b} \left(2 r^2\right)^{a-\frac{1}{2}}}{2\, b! f^{a+b+\frac{1}{2}}} \Bigg\{\sum _{k=1}^b \sum _{j=0}^l \frac{(-1)^{j+1} j! \binom{l}{j} \binom{\frac{1}{2} (l+j-1)}{l}}{2(-k+j+1)!} \frac{\partial ^{b-k}}{\partial x^{b-k}}\left((1-x)^{a+b-\frac{1}{2}} e^{-|\omega|  \sqrt{\frac{2 (1-x) r^2}{f}}}\right)\nonumber\\
    &\quad+\sum _{k=b}^l \sum _{j=0}^{k-b} \frac{k! \binom{l}{k} \binom{\frac{1}{2} (k+l-1)}{l} (-1)^{-2 b+k-j}}{(k-b)!} \binom{k-b}{j} \left(|\omega|  \sqrt{\frac{2 r^2}{f}}\right)^{-2 a-2 k+2 j-1} \Gamma \left(2 a+2 k-2 j+1,0,\frac{2 r |\omega|}{\sqrt{f}} \right)\Bigg\},
\end{align}
\begin{align}
     \mathcal{J}_{\omega l}(a,b\,|r)&=\sqrt{\pi}(2r^{2}\rho^{2})^{a-b}(a-b)!2^{2a}(-1)^{l}\sum_{k=0}^{a-b}\frac{\theta\left(\lambda^{2}-\omega^{2}\right)(a-k)!k!|\omega|^{2k-2a-1}}{(a-k-b)!\Gamma(k-a+\tfrac{1}{2})(k+l+1)!(k-l)!(2\rho^{2})^{k}}\nonumber\\
     &\quad+(2r^2)^{a-b}(-1)^b \Gamma(1+a-b)\frac{r}{f^{1/2}} \sum_{k=0}^{a-b+1}\frac{(-1)^k}{2^{a-b+2}}\binom{a-b+1}{k}\frac{(2l+2(a-b)-4k+3)}{(l-k+\frac12)_{a-b+2}}\nonumber\\ &\quad\times\frac{\partial^{2b}}{\partial|\omega|^{2b}}I_{l+\frac32+a-b-2k}\left(\frac{|\omega|  r}{\sqrt{f}}\right) K_{l+\frac32+a-b-2k}\left(\frac{|\omega|  r}{\sqrt{f}}\right)\nonumber\\
     &=\sqrt{\pi}(2r^{2}\rho^{2})^{a-b}(a-b)!2^{2a}(-1)^{l}\sum_{k=0}^{a-b}\frac{\theta\left(\lambda^{2}-\omega^{2}\right)(a-k)!k!|\omega|^{2k-2a-1}}{(a-k-b)!\Gamma(k-a+\tfrac{1}{2})(k+l+1)!(k-l)!(2\rho^{2})^{k}}\nonumber\\
     &\quad+(2r^2)^{a-b}(-1)^b 
    \left(\frac{r}{\sqrt{f}}\right)^{2b+1}\Gamma(1+a-b)\sum_{k=0}^{a-b+1}\frac{(-1)^k}{2^{a-b+2}}\binom{a-b+1}{k}\frac{(2l+2(a-b)-4k+3)}{(l-k+\frac12)_{a-b+2}}\nonumber\\ 
    &\quad\times\sum_{p=0}^{2b}\binom{2b}{p}\frac{(-1)^p}{2^{2b}}\sum_{q=0}^{4b-2p}\binom{2b-p}{q}I_{l+3/2+a-3b-2k+p+2q}\left(\frac{|\omega|  r}{\sqrt{f}}\right)\sum_{n=0}^{2p}\binom{p}{n}K_{l+3/2+a-b-2k-p+2n}\left(\frac{|\omega|  r}{\sqrt{f}}\right)\nonumber\\
    &=-\lim_{|\omega'|\to|\omega|}\frac{f^{a-b}}{4 \pi } \sum _{k=0}^{a-b} 2^{l+2} (-1)^{a-k} \binom{a-b}{k} \frac{\partial ^{2 a-2 k}}{\partial |\omega'|^{2 a-2 k}}\sum _{n=0}^l \sum _{b=0}^n \frac{\pi}{|\omega'|}  (-1)^{n-b} \binom{l}{n} \binom{\frac{1}{2} (l+n-1)}{l} \binom{n}{b}\nonumber\\
    &\quad\times\left(\frac{2 r^2}{f}\right)^k \left(\frac{\Gamma \left(2 (k-b+n+1),0, \frac{2 |\omega'| r}{\sqrt{f}}\right)}{\left(\frac{\sqrt{2} |\omega'| r}{\sqrt{f}}\right)^{2 (k-b+n+1)}}-\frac{2^{k-b+n} \theta (\lambda^2-\omega^2 )}{k-b+n+1}\right).
\end{align}

\section{Hadamard Coefficients}
\label{app:HCoeff}
Below we list the coefficients $\mathcal{D}_{ab}^{(\mathrm{p})}$, $\mathcal{D}_{ab}^{(\mathrm{r})}$, $\mathcal{T}_{ab}^{(\mathrm{l})}$, $\mathcal{T}_{ab}^{(\mathrm{r})}$ and $\mathcal{T}_{ab}^{(\mathrm{p})}$  to 2nd order.

 \begin{align*}
 	&\mathcal{D}_{00}^{(\mathrm{r})}=2,\nonumber\\
 	&\mathcal{D}_{10}^{(\mathrm{r})}=-\frac{f \left(r^2 f''-2 r f'+2 f-2\right)}{12 r^2},\nonumber\\
 	&\mathcal{D}_{11}^{(\mathrm{r})}=\frac{f \left(r^2 \left(f'^2\right)-4 f \left(r f'+1\right)+4
 		f^2\right)}{24 r^2},\nonumber\\
 	&\mathcal{D}_{20}^{(\mathrm{r})}=\frac{1}{2880 r^4}\Bigg(f \Big[-5 r^2 \left(-f'^2\right) \left(r^2 f''-2 r
 	f'-2\right)-8 f^2 \left(3 r^3 f^{(3)}-7 r^2 f''+19 r
 	f'+10\right)\nonumber\\
 	&\qquad\quad+f \left(9 r^4 f''^2-20 r^2 f''+86 r^2 f'^2+4 r f'
 	\left(3 r^3 f^{(3)}-14 r^2 f''+20\right)+4\right)+76
 	f^3\Big]\Bigg),\nonumber\\
 	&\mathcal{D}_{21}^{(\mathrm{r})}=-\frac{1}{2880 r^4}\Bigg(f \Big[r^4 \Big( f'^2+f'^4\Big)+r^2 f
 	\Big(-30 r
 	f'^3+f'^2 \Big(11 r^2 f''-30\Big)\Big)+4 f^3 \Big(11 r^2 f''-52
 	r f'-40\Big)\nonumber\\
 	&\qquad\qquad\qquad\qquad\quad-2 f^2 \Big(10 r^2 f''-67 r^2 f'^2 +f' \Big(22 r^3
 	f''-80 r\Big) -28\Big)+104 f^4\Big]\Bigg),\nonumber\\
 	&\mathcal{D}_{22}^{(\mathrm{r})}=\frac{f^2 \left(r^2 f'^2-4 f \left(r f'+1\right)+4
 		f^2\right)^2}{1152 r^4},\nonumber\\
 	&\mathcal{D}_{11}^{(\mathrm{p})}=-\frac{f'}{6 r},\nonumber\\
 	&\mathcal{D}_{21}^{(\mathrm{p})}=\frac{f \left(-9 r f'^2+6 r f \left(r f^{(3)}-2 f''\right)+2 f'
 		\left(7 r^2 f''+f+5\right)\right)}{720 r^3},\nonumber\\
 	&\mathcal{D}_{22}^{(\mathrm{p})}=\frac{7 r^2 f'^2-10 r f'+r f \left(9 r f''+4 f'\right)-3 f^2+3}{720
 		r^4},\nonumber\\
 	&\mathcal{T}_{00}^{(\mathrm{l})}=\frac{6 \mu ^2 r^2-(6 \xi -1) \left(r^2 f''+4 r f'+2 f-2\right)}{12 r^2},\nonumber\\
 	&\mathcal{T}_{10}^{(\mathrm{l})}=\frac{1}{480 r^4}\Bigg(r^2 f'' \left((10 \xi  (3 \xi -1)+1) r^2 f''-10 (6 \xi -1) \left(2 \xi +\mu ^2 r^2\right)\right)+4 (5 \xi  (24 \xi -5)+1) r^2 f'^2\nonumber\\
 	&\qquad\qquad+2 r f' \left((5 \xi -1) r^3 f^{(3)}+2 \left(60 \xi ^2-5 \xi -1\right) r^2 f''+40 (1-6 \xi ) \xi
 	+10 \mu ^2 (1-12 \xi ) r^2\right)\nonumber\\
 	&	\qquad\qquad+2r f \left((80 \xi  (3 \xi -1)+6) f'+r \left((20 \xi  (3 \xi +2)-8) f''+r \left((5 \xi -1) r f^{(4)}(r)+(40 \xi -7) f^{(3)}\right)\right)\right)\nonumber\\ 
 	&	\qquad\qquad\qquad\qquad\qquad\qquad\qquad\qquad-40 f \xi  \left(6 \xi +3 \mu ^2	r^2-1\right)+4 (10 \xi  (3 \xi -1)+1) f^2+2 \left(15 \left(2 \xi +\mu ^2 r^2\right)^2-2\right)\Bigg),\nonumber\\
 	&\mathcal{T}_{11}^{(\mathrm{l})}=\frac{f}{480 r^4}\Bigg(-2 \left(r^2 f''+2\right) \left((1-5 \xi ) r^2 f''+10 \xi +5 \mu ^2 r^2-2\right)+2 (1-10 \xi ) r^2 f'^2\nonumber\\
 	&\qquad\qquad+r f \left((16-80 \xi ) f'+r \left((12-80 \xi )
 	f''+r \left(r f^{(4)}(r)+(6-20 \xi ) f^{(3)}\right)+20 \mu ^2\right)\right)+8 (5 \xi -1) f^2\nonumber\\
 	&\qquad\qquad\qquad\qquad\qquad\qquad\qquad\qquad\qquad\qquad\qquad\qquad+r f' \left((10 \xi -1) r^3 f^{(3)}+4 (20 \xi -3) r^2 f''+20 (6 \xi -1)\right)\Bigg)\nonumber\\
 	&\mathcal{T}_{10}^{(\mathrm{r})}=\frac{f \left(r^2 f'^2-4 f \left(r f'+1\right)+4
 		f^2\right) \left((1-6 \xi) \left(2-r \left(r f''+4 f'\right)-2 f+2\right)-6 \mu ^2 r^2\right)}{576 r^4},\nonumber\\
 	&\mathcal{T}_{10}^{(\mathrm{p})}=\frac{(f-1) \left((1-6 \xi ) \left(2-r \left(r f''+4 f'\right)-2 f\right)-6 \mu ^2 r^2\right)}{144 r^4},\nonumber\\
 	&\mathcal{T}_{11}^{(\mathrm{p})}=\frac{f \left(r f'-2 f+2\right) \left((1-6 \xi) \left(2-r \left(r f''+4 f'\right)-2 f\right)-6 \mu ^2 r^2\right)}{144 r^4}.\nonumber\\
 \end{align*}
 
\section{Analytical Part of the RSET}
\label{app:TAnalytical}
In this appendix, we give explicit expressions for $\langle \hat{T}^{\mu}{}_{\nu}\rangle_{\textrm{analytical}}$ that arises in the split in Eq.~(\ref{eq:Tsplit}) of the RSET into a numerical and analytical part. The analytical part is separately conserved and given by
\begin{align}
   \langle \hat{T}^{\mu}{}_{\nu}\rangle_{\textrm{analytical}}=-\bar{\mathsf{w}}^{\mu}{}_{\nu}-(\xi-\tfrac{1}{2})\bar{\mathsf{w}}^{;\mu}{}_{\nu}+(\xi-\tfrac{1}{2})\Box\bar{\mathsf{w}} +\xi\,R^{\mu}{}_{\nu}\bar{\mathsf{w}}-\frac{1}{8\pi^{2}}v_{1}, 
\end{align}
where
\begin{align}
    \bar{\mathsf{w}}&=-\frac{1}{8\pi^{2}}\left\{\frac{\mathcal{D}_{1 0}^{(\mathrm{r})}}{f}+\frac{\mathcal{D}_{2 2}^{(\mathrm{r})}}{f^{2}}+\mathcal{D}_{1 1}^{(\mathrm{p})}+\mathcal{T}_{00}^{(\textrm{l})}(r)(\log( f /(\ell^{2}\lambda^2))-2\gamma_{\textrm{E}})\right\}\\
    \bar{\mathsf{w}}^{\tau}{}_{\tau}&=\frac{1}{4}f'\,\bar{\mathsf{w}}'-\frac{1}{4\pi^{2}}\Bigg\{\frac{\mathcal{T}_{1 0}^{(\mathrm{r})}}{f^{2}}+\mathcal{T}_{1 0}^{(\mathrm{p})}+\frac{\mathcal{T}_{1 1}^{(\mathrm{p})}}{f}+\frac{\mathcal{T}_{1 0}^{(\mathrm{r})}}{f^{2}}+\mathcal{D}_{22}^{(\mathrm{p})}+\frac{\mathcal{D}_{21}^{(\mathrm{p})}}{f}+\frac{\mathcal{D}_{20}^{(\mathrm{r})}}{f^{2}}+\frac{\mathcal{D}_{21}^{(\mathrm{r})}}{f^{3}}+\frac{\mathcal{D}_{22}^{(\mathrm{r})}}{f^{4}}\nonumber\\
    &\qquad \qquad\qquad+(f\mathcal{T}_{10}^{(\textrm{l})}+\mathcal{T}_{11}^{(\textrm{l})})(\log( f /(\ell^{2}\lambda^{2}))-2\gamma_{\textrm{E}}+3)\Bigg\}\\
    \bar{\mathsf{w}}^{\phi}{}_{\phi}&=\bar{\mathsf{w}}^{\theta}{}_{\theta}=\frac{1}{2r}f\,\bar{\mathsf{w}}'-\frac{1}{4\pi^{2}}\Bigg\{\mathcal{T}_{1 0}^{(\mathrm{p})}+\frac{\mathcal{T}_{1 1}^{(\mathrm{l})}}{f}-\frac{\mathcal{T}_{1 0}^{(\mathrm{r})}}{f^{2}}+\mathcal{D}_{22}^{(\mathrm{p})}-\frac{\mathcal{D}_{20}^{(\mathrm{r})}}{f^{2}}-2\frac{\mathcal{D}_{21}^{(\mathrm{r})}}{f^{3}}-3\frac{\mathcal{D}_{22}^{(\mathrm{r})}}{f^{4}}\nonumber\\
    &\qquad\qquad\qquad+r^2\mathcal{T}_{10}^{(\textrm{l})}(\log( f /(\ell^{2}\lambda^{2}))-2\gamma_{\textrm{E}}+1)\Bigg\}\\
    \bar{\mathsf{w}}^{r}{}_{r}&=-\bar{\mathsf{w}}^{\tau}{}_{\tau}-2\bar{\mathsf{w}}^{\phi}{}_{\phi}+\xi\,R\,\bar{\mathsf{w}}+\mu^{2}\bar{\mathsf{w}}-\frac{3}{4\pi^{2}}v_{1},
\end{align}
where explicit expressions for the Hadamard coefficients in terms of $f$ are found in the previous appendix.
\end{widetext}

\bibliographystyle{apsrev4-1}
	\bibliography{bib}
\end{document}